\documentclass[conference]{IEEEtran}

\usepackage[normalem]{ulem}

\usepackage{amsmath,amssymb,amsfonts}
\usepackage{xcolor}
\usepackage{rotating}
\usepackage{algorithm}
\usepackage{slantsc}
\usepackage{comment}
\usepackage{multicol}
\usepackage{multirow}
\usepackage{soul}
\usepackage{color, colortbl}
\usepackage{arydshln}
\usepackage{xspace}
\definecolor{Gray}{gray}{0.9}
\definecolor{LightCyan}{rgb}{0.88,1,1}

\newcolumntype{a}{>{\columncolor{Gray}}l}
\newcolumntype{d}{>{\columncolor{LightCyan}}l}

\usepackage{algpseudocode}
\newcommand{\fixme}[1]{\textbf{***FIXME: #1***}}
\usepackage{tikz}
\newcommand*\circled[1]{\tikz[baseline=(char.base)]{\node[fill=gray,shape=circle,draw,inner sep=0.6pt] (char) {#1};}}

\newcommand{\tname}{{\sc DNNShield}\xspace}

\title{
DNNShield: Dynamic Randomized Model Sparsification, A Defense Against Adversarial Machine Learning}

\author{
    \IEEEauthorblockN{Mohammad Hossein Samavatian, Saikat Majumdar, Kristin Barber, Radu Teodorescu}
    \IEEEauthorblockA{Department of Computer Science and Engineering \\ The Ohio State University, Columbus, OH, USA
    \\ \{samavatian.1, majumdar.42, barber.245, teodorescu.1\}@osu.edu}
}

\begin{document}

\maketitle
\pagestyle{plain}



\begin{abstract}

DNNs are known to be vulnerable to so-called adversarial attacks that manipulate inputs to cause incorrect results that can be beneficial to an attacker or damaging to the victim. Recent works have proposed approximate computation as a defense mechanism against machine learning attacks. We show that these approaches, while successful for a range of inputs, are insufficient to address stronger, high-confidence adversarial attacks. To address this, we propose \tname, a hardware-accelerated defense that adapts the strength of the response to the confidence of the adversarial input. 
Our approach relies on \underline{dynamic} and \underline{random} sparsification of the DNN model to achieve inference approximation efficiently and with fine-grain control over the approximation error. \tname uses the output distribution characteristics of sparsified inference compared to a dense reference to detect adversarial inputs. We show an adversarial detection rate of 86\% when applied to VGG16 and 88\% when applied to ResNet50, which exceeds the detection rate of the state of the art approaches, with a much lower overhead. We demonstrate a software/hardware-accelerated FPGA prototype, which reduces the performance impact of \tname relative to software-only CPU and GPU implementations. 

\end{abstract}


\section{Introduction}
\label{sec:introduction}

Deep neural networks (DNNs) are rapidly becoming indispensable tools for solving an increasingly diverse set of complex problems, including computer vision,
 natural language processing,
 machine translation,
 and many others. 
Some of these application domains, such as medical, self-driving cars, face recognition, etc. expect high robustness in order to gain public trust and achieve commercial adoption. Unfortunately, DNNs are known to be vulnerable to so-called ``adversarial attacks'' that purposefully compel classification algorithms to produce erroneous results. For example, in the computer vision domain, a large number of attacks \cite{carlini2017towards, moosavi-dezfooli2017universal, moosavi-dezfooli2016deepfool, goodfellow2015explaining, chen2018ead, kurakin2017adversarial, papernot2016the, madry2018towards} have demonstrated the ability to force state-of-the art classifiers such as Inception\cite{szegedy2016rethinking}, VGG\cite{simonyan2014deep}, ResNet\cite{ResNet}, etc. to misclassify inputs that are carefully manipulated by an attacker. In most attacks, input images are only slightly altered such that they appear to the casual observer to be unchanged. However the alterations are made with sophisticated algorithms which, in spite of the imperceptible changes to the input, result in reliable misclassification.

Several defenses have been proposed to address adversarial attacks \cite{ma2019nic,papernot2016distillation,xu2018feature,dhillon2018stochastic,cao2017mitigating,ma2018characterizing}. Most rely on purely software implementations, with high overheads, limiting their utility to real-world applications.  A recent line of research has explored hardware-assisted approximate computing to introduce controlled errors into the inference process, either through model quantization \cite{fu2021,panda2020quanos} or approximate computation \cite{guesmi2020defensive}. This inference approximation disrupts the effect of the adversarial modifications, making the attacks less likely to succeed. At the same time, if the errors are kept small, approximate inference tends to have less effect on benign inputs' classification accuracy. 




We investigate the scalability of defensive approximation approaches to a broader class of attacks. We find that, while approximation methods work well for some inputs, they do not scale well to strong adversarial attacks that are trained to have high classification confidence. This is because the noise introduced through approximation is insufficient to reverse the adversarial effects. We also show that, even if noise is increased, full recovery of strong adversarials is less likely. We therefore argue that defensive techniques should focus on detecting adversarial inputs, which has higher probability of success, rather than recovery of the original class.  

This paper presents \tname, a hardware/software co-designed defense that takes a different approach to inference approximation and addresses some of the limitations of prior approaches. \tname is an online adversarial detection framework that uses the effects of model sparsification to discriminate between adversarial and benign inputs. \tname runs both precise and sparse inference passes for each input and compares their output. It then uses the deviation in output classification that is triggered by the sparsification to classify the inputs as benign or adversarial. 

A key observation we make in this work is that {\bf tailoring the approximation error rate to the confidence of the input classification dramatically increases the adversarial detection rate}, while at the same time maintaining a low false positive rate for benign inputs. \tname is the first work to recognize the importance of this correlation for accurate adversarial detection. Unlike prior work, \tname \textbf{dynamically} and \textbf{randomly} varies the approximation error and distribution. Dynamic approximation error is needed to adapt to the confidence of diverse inputs. Randomness in the error distribution is crucial in ensuring that adversaries cannot re-train to account for predictable inference noise.  

To achieve these goals \tname uses hardware-assisted dynamic and random model sparsification to implement approximate inference. Model sparsification involves dropping weights from the model, and has been used to improve performance and energy efficiency \cite{deng2021gospa,gong2020save}. \tname controls the sparsification rate dynamically to enable flexible control over the approximation error. Sparsification is also random to make the noise input independent and consequently training defense-aware attacks difficult. 

\tname demonstrates robust detection across a broad set of attacks, with high accuracy and low false positive rate. We show an adversarial detection rate of 86\% when applied to VGG16 and 88\% when applied to ResNet50, which exceeds the detection rate of the state of the art approaches. We also show that \tname is robust against attacks that are aware of our defense and attempt to circumvent it. 





\tname requires multiple inference passes, increasing inference latency. To mitigate this overhead we propose a hardware/software co-designed accelerator aimed at reducing the performance overhead. The accelerator design builds explicit support for dynamic and random model sparsification. The \tname accelerator is optimized for efficiently executing sparsified models in which the sparsification rate changes as a function of the input -- which is more challenging compared to models for which weight sparsity is fixed. We show that the \tname accelerator reduces the performance impact of approximate inference-based adversarial detection to $1.53\times-2\times$ relative to the unprotected baseline, compared to $15\times$--$25\times$ overhead for a software-only GPU implementation.

This paper makes the following contributions: 
\begin{itemize}


\item First work to use DNN sparsification as a dynamic and random approximation-based defense against machine learning attacks. 

\item First adaptive defense that adjusts the approximation error in correlation with the confidence of the classification, increasing detection accuracy even for strong attacks. 


\item Presents \tname, a software/hardware-accelerated co-design of the proposed defense and demonstrates its robustness against defense-aware attacks. 

\item Evaluates \tname through a proof-of-concept implementation in the Xilinx CHaiDNN accelerator, showing $<2.65\%$ area overhead and $<5.1\%$ power overhead.

\end{itemize}

\section{Background - Adversarial Attacks}
\label{sec:background}


Adversarial attacks were first introduced by Szegedy et al. in \cite{szegedy2014intriguing}.
The objective of an adversarial attack is to force the output classification for some maliciously crafted input \textbf{x$^\prime$}, based on a benign input \textbf{x}.  Attacks can be \emph{targeted}, where the adversary's goal is for \textbf{x$^\prime$} to be misclassified as a particular class \emph{t}, or \emph{untargeted}, in which misclassification of \textbf{x$^\prime$} to any other class is sufficient. Targeted attacks were formally solving the following optimization problem \cite{szegedy2014intriguing}:
\begin{equation} \label{eq1}
\textrm{min }\;d(x,x + \delta) \\
\textrm{subject to: }\;C(x + \delta) = t,\\
x + \delta \in [0,1]^n,
\end{equation}
where $\delta$ is the added noise, \emph{t} is the desired target label for the adversarial example produced by $x + \delta$, and \emph{d} is a metric to measure the distortion from benign example and the adversarial one. 
$L_p$ norm is widely used as distortion metric. In this paper we refer to the $L_2$-norm as the distortion metric which is $L_2 = \sqrt{\sum (x'-x)^2} $. 

In this work we focus on two strong state of the art attacks. The {\bf Carlini-Wagner (CW)} attack \cite{carlini2017towards} has been shown to be very effective at creating adversarial images that looks very similar to original inputs and is successful at attacking models protected by defensive methods such as defensive distillation \cite{papernot2016distillation}. The CW attack uses the loss function:
\begin{equation}
    loss(x') = max(max\{Z(x')_i : i \neq t\} - Z(x')_t, - k)
\end{equation}
where $Z$ is the logit of classifier and $k$ is controlling the confidence with which the misclassification occurs.

The \textbf{EAD} attack \cite{chen2018ead} generalizes the $CW$  attack by performing elastic-net regularization, linearly combining the $L_1$ and $L_2$ penalty functions. The EAD attack attains superior performance in transfer attacks which can reduce the effectiveness of defenses that are based on attack transfer such as \cite{pmlr_v139_fu21c,fu2021}. The EAD attack has also defeated several prior defenses \cite{lu2018limitation,sharma2018bypassing,sharma2017attacking,sharma2017attacking_madry},

\section{Motivation}

Strong attacks such as CW and EAD can be tuned to produce a class of adversarial inputs that present a significant challenge to approximation-based defenses. Prior work has shown that adversarial inputs can be constructed to induce misclassification with very high confidence \cite{sharma2018bypassing,carlini2019evaluating,carlini2017towards}. In other words, the victim model assigns a very high probability to the adversarial input belonging to the wrong class. These so-called ``high confidence'' adversarials can be constructed while minimizing the distortion to original input.

\subsection{High-Confidence Attack Variants}
\begin{figure}[htb]
\centering
 \includegraphics[width=\linewidth]{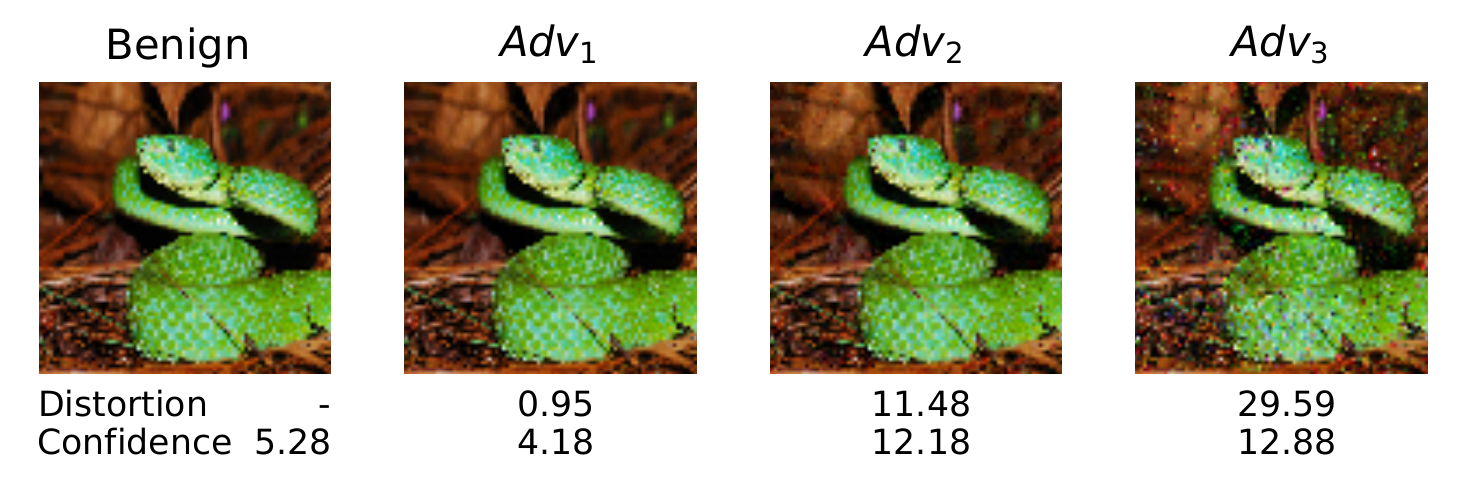}
\caption{Benign and adversarial examples with various distortion and confidence values.}\label{Fig:stronge_adv}
\end{figure}

Figure \ref{Fig:stronge_adv} shows an example of multiple adversarial samples for a benign image with different levels of classification confidence and the corresponding distortion. To measure classification confidence we used the Z-score (the number of standard deviations by which the value of a raw score is above or below the mean value) of the maximum logit value, which corresponds to the class with the highest confidence.  $Adv_1$ is a low distortion adversarial of the benign with low classification confidence of 4.18. $Adv_2$ is a high confidence example of the same input with very high classification confidence of 12. While distortion is also higher, it is still imperceptible to the naked eye. We will show that existing defenses are ineffective against this type of adversarial. Increasing the confidence beyond 12 increased distortion significantly, as $Adv_3$ shows. 

\subsection{Classification Confidence and Approximation}
\begin{figure}[!htb]
\minipage{\linewidth}
  \includegraphics[width=\linewidth]{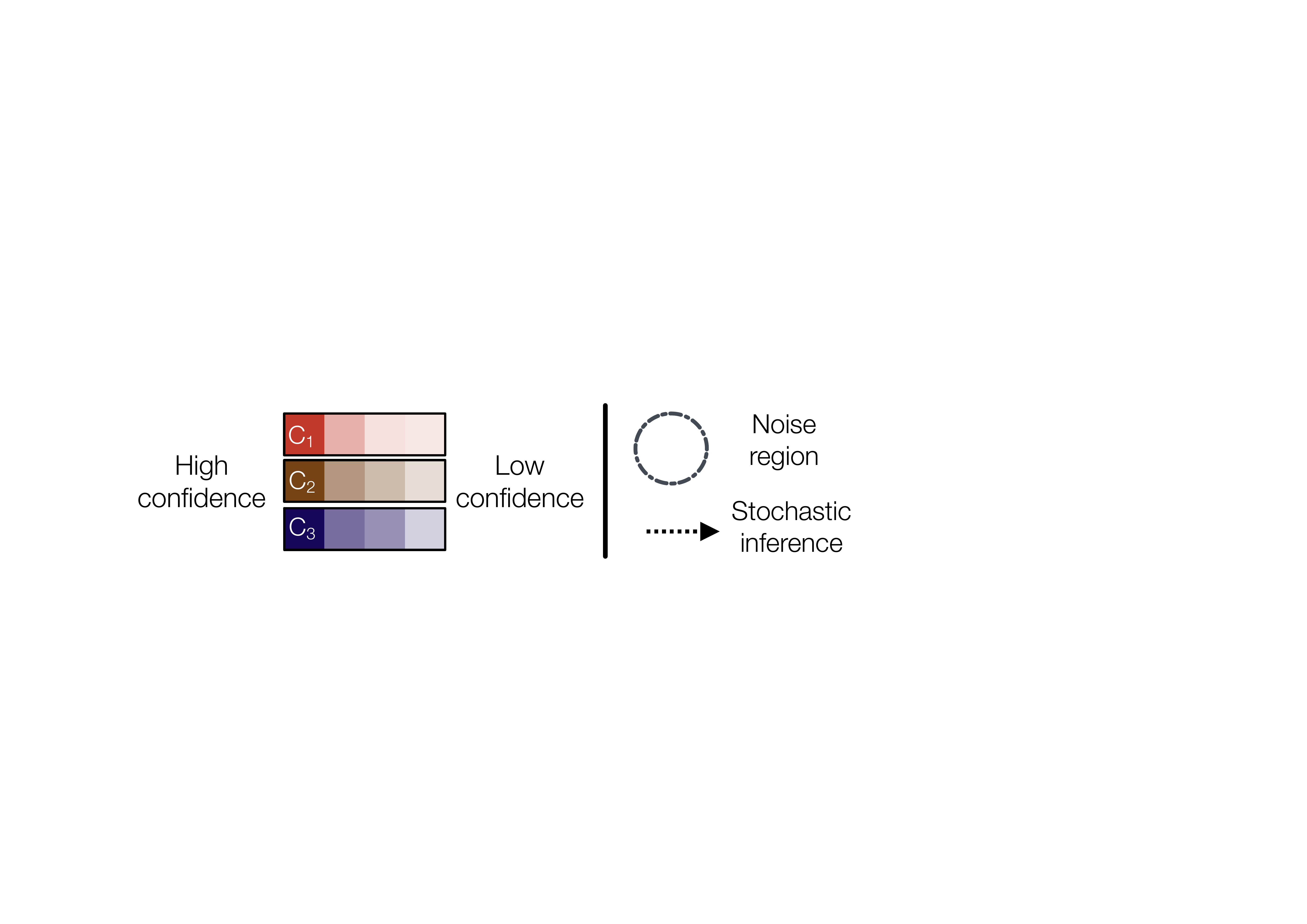}
\endminipage\hfill
\minipage{0.32\linewidth}
  \includegraphics[width=\linewidth]{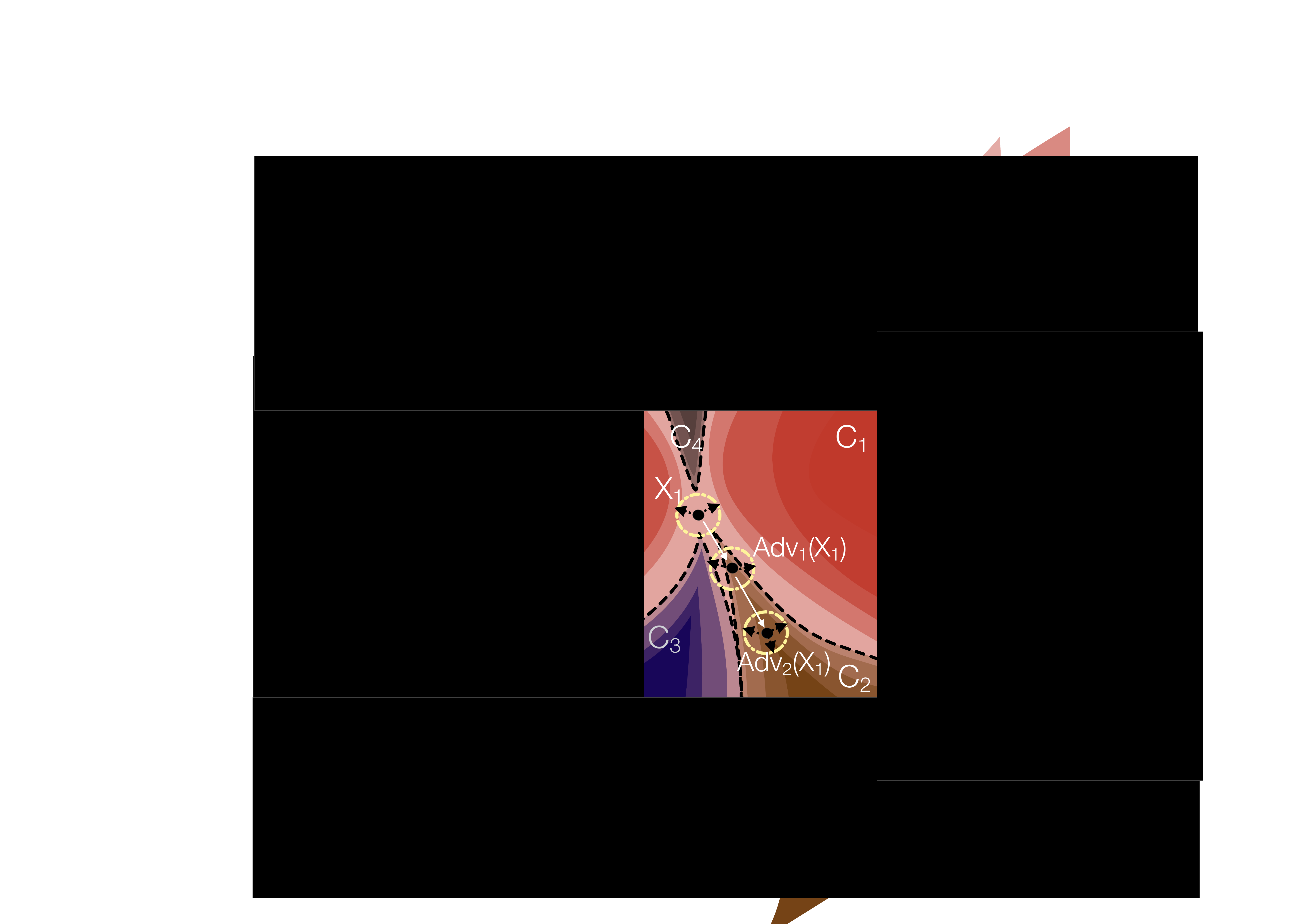}
  {\small Low noise}\label{fig:1a}
\endminipage\hfill
\minipage{0.32\linewidth}
  \includegraphics[width=\linewidth]{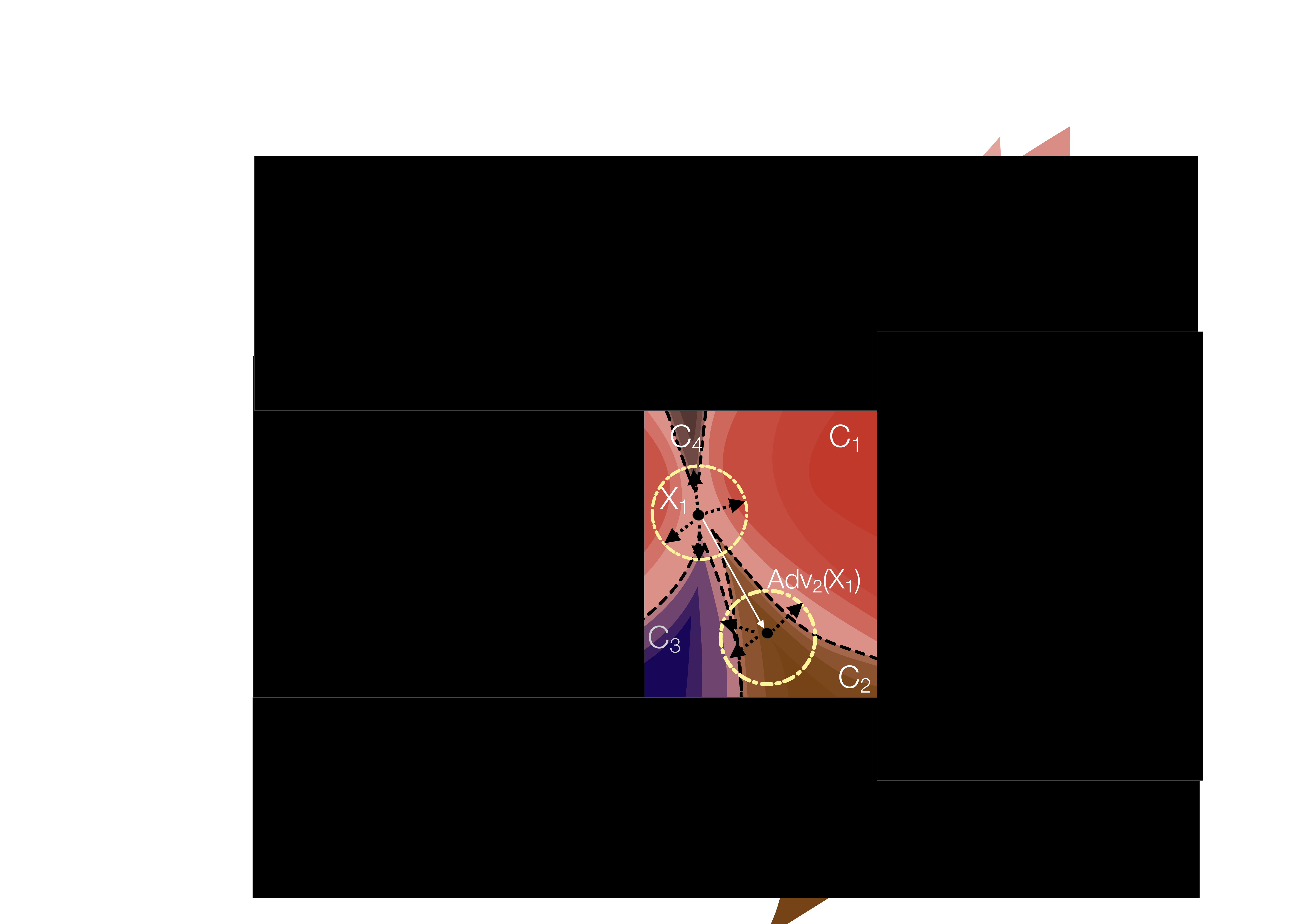}
  {\small High noise}\label{fig:1b}
\endminipage\hfill
\minipage{0.32\linewidth}%
  \includegraphics[width=\linewidth]{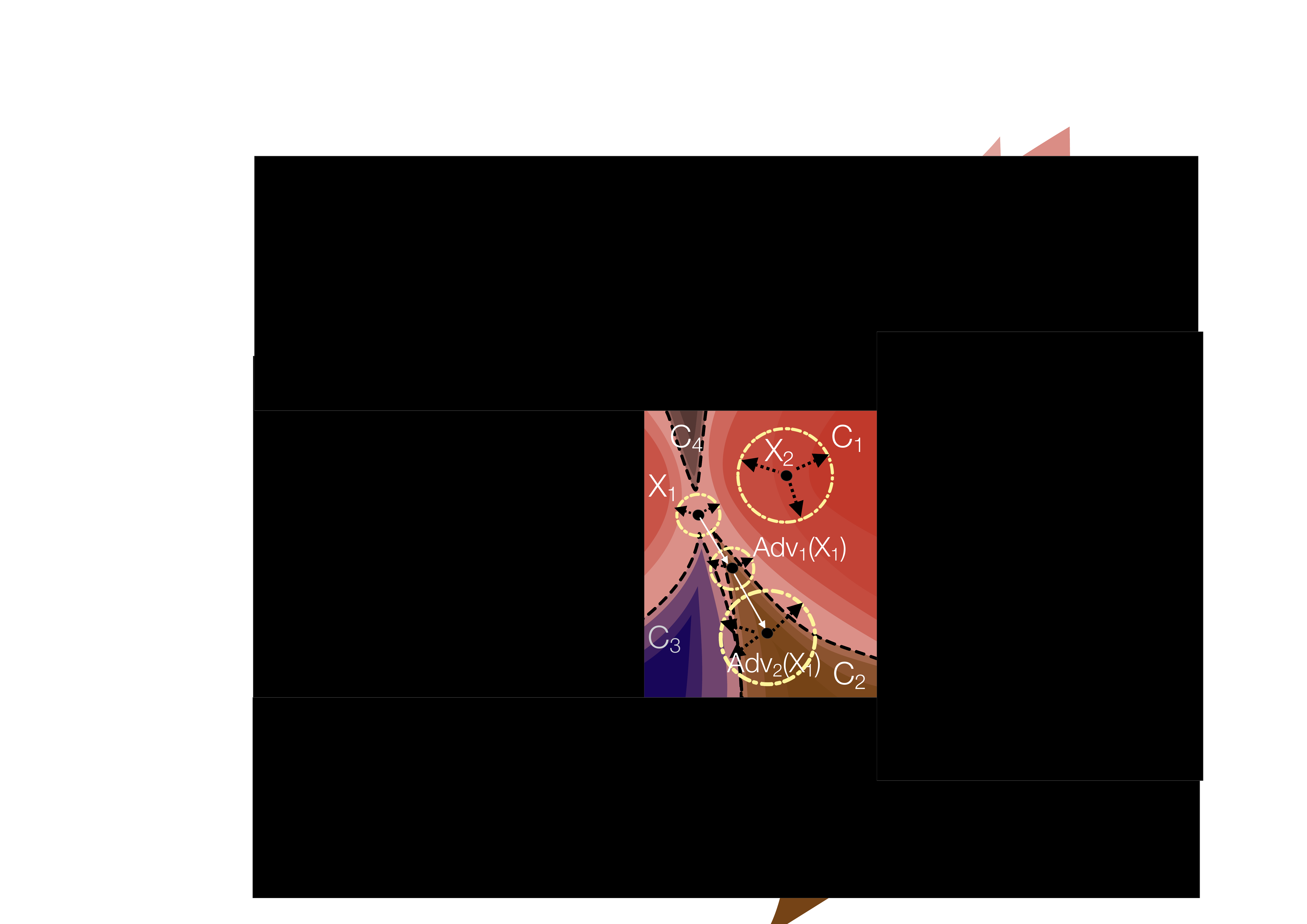}
  {\small Adaptive noise}\label{fig:1c}
\endminipage
\caption{Schematic illustration of decision regions of a base classifier. Different classes are drawn in different colors with darker shades indicating higher confidence.  $X_1$ and $X_2$ are benign inputs, $Adv_1(X_1)$ and $Adv_2(X_1)$ are low confidence and high confidence adversarial inputs generated from $X_1$. The circles shows range of classification deviation under a certain amount of noise.}\label{Fig:decision_region}
\end{figure}

Approximate computing defense methods introduce noise into the input and/or the model and often result in the recovery of the original class. Figure \ref{Fig:decision_region} schematically illustrates how approximation can recover adversarial inputs. Figure \ref{Fig:decision_region} illustrates the decision space of a classifier, with four output classes: $C_1$, $C_2$, $C_3$ and $C_4$, represented as regions with different colors. The darkness of the color represents the confidence of the classification. We consider a benign input $X_1$ classified with low in class $C_1$, and an adversarial sample $Adv_{1}(X_1)$ that is misclassified into class $C_2$ with low confidence. The dotted circles around each input represent the range of classification outcomes as a result of noise. 


Prior work has observed that high-quality adversarial inputs occur with low probability, which means they reside in small and low-density pockets of the classification space. As a result, their output class distribution differs from that of their closest data submanifold \cite{feinman2017detecting}. We can see in Figure \ref{Fig:decision_region} (a) that $Adv_{1}(X_1)$ resides in a narrow cone of class $C_2$, where benign images do not generally exist. This means that approximation can easily change the classification of $Adv_{1}(X_1)$ and move it back into its original class, $C_1$. 

In this work we show for the first time, that high-confidence adversarials do not respond in the same way to approximation errors. To illustrate why, let's consider $Adv_{2}(X_1)$, a high-confidence adversarial of $X_1$. As Figure \ref{Fig:decision_region} (a) shows, for the same error, the classification area of $Adv_{2}(X_1)$ lands within region $C_2$, failing to recover the original class $C_1$. 

Figure \ref{Fig:decision_region} (b) shows that, if we increase the approximation error, the probability of recovering $Adv_{2}(X_1)$ increases. However, if the same approximation error is applied uniformly to $X_1$, there is an increased probability that $X_1$ will be misclassified, resulting in false positives. 

In order to address this limitation, we propose {\em correlating the approximation error to the confidence of the classification}. Figure \ref{Fig:decision_region} (c) illustrates this with circles of different radii: small radius corresponding to lower approximation error for $X_1$ and $Adv_{1}(X_1)$ -- which are low confidence classifications -- and larger radius (approximation error) for $Adv_{1}(X_1)$ and a high confidence benign, $X_2$. We can see that the low error does not lead to misclassification for $X_1$, while recovering $Adv_{1}(X_1)$ with high probability. At the same time, a high error rate will not lead to misclassification for $X_2$, while increasing the probability of recovering $Adv_{2}(X_1)$.  

\subsection{Approximation-Based Defenses}

Prior work has used input noise and approximate inference to improve model robustness against adversarial attacks. For example, \cite{hu2019a} and \cite{cohen2019certified} have shown that adding some amount of random noise to images can help DNNs correctly classify adversarial inputs. 

Recent work has proposed using hardware-based approximation methods as similar defenses. Guesmi et al.~\cite{guesmi2020defensive} proposed ``Defensive Approximation" (\textit{DA}) which used custom approximate multipliers, to introduce controlled errors into a CNN accelerator. Similarly, Fu et al.~\cite{fu2021} used hardware-assisted parameter quantization as the approximation mechanism. Model quantization is the process of reducing the precision of the model parameter representation, and has been used to improve performance, energy and storage efficiency of DNNs. In~\cite{fu2021} a \textit{2-in-1} hardware accelerator dynamically chooses between 12 quantization levels to use at inference, introducing approximation into the model. While these approaches are effective and have low overheads, they use either fixed approximation error~\cite{guesmi2020defensive} or randomly-selected error from a limited set of up to 12 precision levels~\cite{fu2021}. In addition, both techniques generate input-dependent noise, which an attacker could reproduce to circumvent the defense. 

In order to study the response of these approximation techniques to high confidence adversarials, we re-implemented both the \textit{DA} and the \textit{2-in-1} defenses for VGG16 and ResNet50 models. The original DA defense uses a single AMA5 floating-point based approximate multiplier. In order to explore a wider range of approximation errors we use a set of seven Int8 approximate multipliers from the \textit{Evoapproxlib} library \cite{evoapprox16}. We created approximated models using approximate convolution (\textit{AxConv}) implementations from \cite{Vaverka_DATE,Mrazek_ICCAD}. We also used the quantized model from \textit{2-in-1}~\cite{fu2021} and generated high confidence adversarial samples targeting two fixed quantization levels of 16 and 4-bit precision. We then tested the 2-in-1 defense on these samples. 
We generated several sets of adversarial samples with different classification confidences by changing the $k$ parameter in the CW and EAD attacks. 

\begin{figure}[tb]
\centering
 \includegraphics[width=0.9\linewidth]{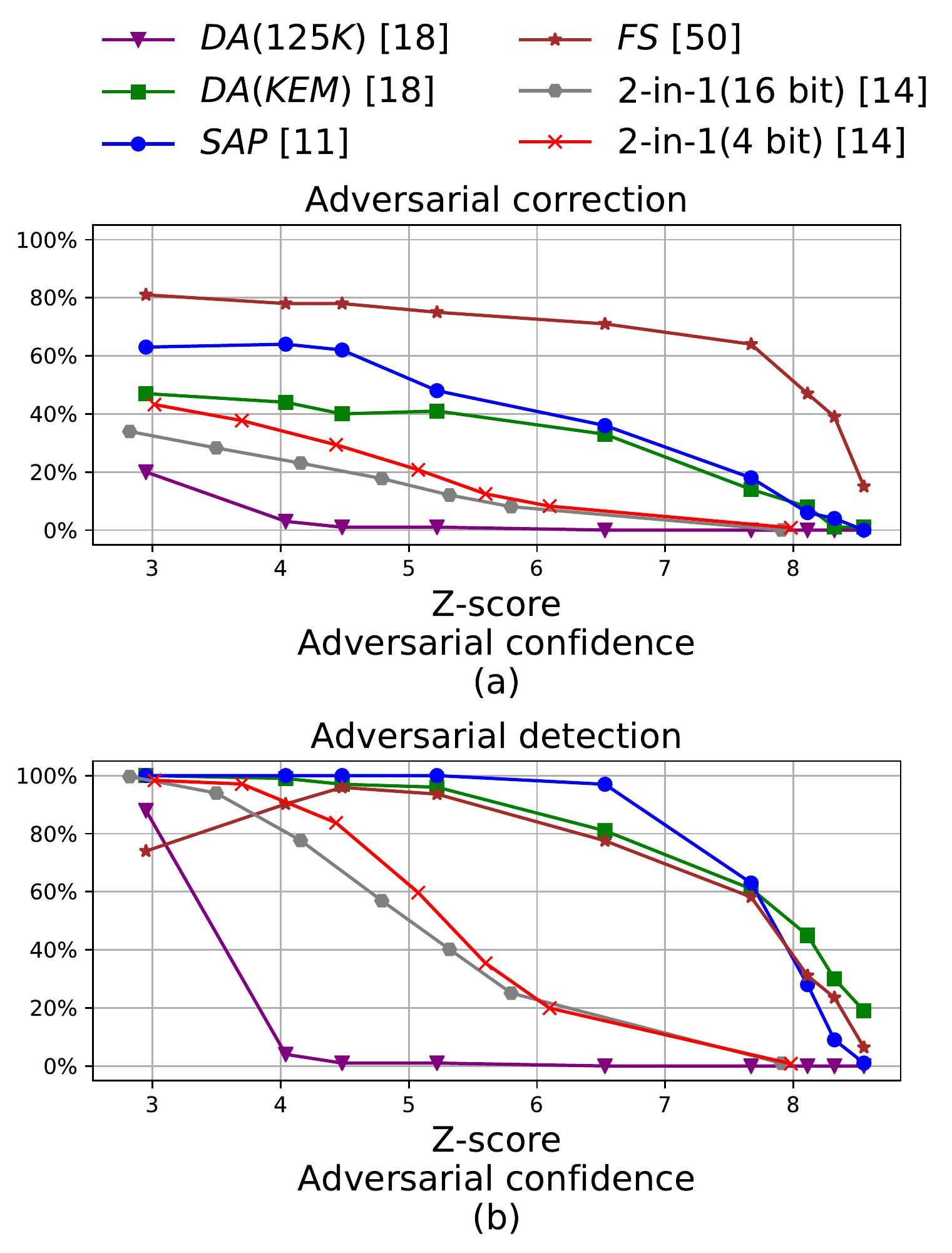}
\caption{(a) Adversarial correction and (b) adversarial detection for different defense methods versus confidence of adversarial attacks on VGG16 and ResNet50.}\label{Fig:prior_works}
\end{figure}

Figure \ref{Fig:prior_works} (a) shows correction rate (percentage of adversarial examples that are correctly classified by the approximate inference) vs. adversarial confidence for two approximate multiplier versions \textit{DA(125K)} and \textit{DA(KEM)}, and the two \textit{2-in-1} adversarial variants. For reference, we also include two software-only approximation techniques: Feature Squeezing (\textit{FS})~\cite{sharma2018bypassing} and \textit{SAP}~\cite{dhillon2018stochastic}. We can see that most approximation techniques perform relatively well with low and medium confidence samples. However, as adversarial confidence increases, correction rate drops below 20\%. The software-only methods perform the best, but they also have the highest overhead. FS, which exhibits the highest correction rate, also has a 4$\times$ performance/energy overhead.

Given the low correction rate of defense methods for high confidence adversarial inputs, we also evaluated an adversarial detection approach. Except $FS$ which has its own detection methodology, for detection we simply compared the classification outputs of the unprotected and protected models for the same input. If the outputs do not match, the input is classified as adversarial. The intuition behind this approach is that the classification of adversarial examples is more likely to change during approximate inference, although the output classification may not be the correct one. This is especially important for very strong adversarial examples which are far from decision boundaries (e.g. $Adv_{2}(X_1)$ in Figure \ref{Fig:decision_region}), and approximation is unlikely to recover the correct classification, as observed in \cite{hosseini2019odds}. However, as we will show in this paper, approximate inference is sufficient to change output distribution of all adversarials in a way that makes them detectable with high probability.

Figure \ref{Fig:prior_works} (b) shows adversarial detection for the same adversarial inputs and defense methods. We can see that the detection rate is much higher than correction for low-confidence adversarials, but still drops below 50\% for adversarials with confidence greater than 8. $FS$ performs better than other methods, but it is still inadequate for high confidence samples. We will compare \tname with FS in our evaluation.

These results show that approximation methods used in prior work, which use fixed error rates are insufficient to detect high confidence adversarial examples.

\subsection{The Case for Adaptive Approximation}

\begin{figure}[tb]
\centering
 \includegraphics[width=0.8\linewidth]{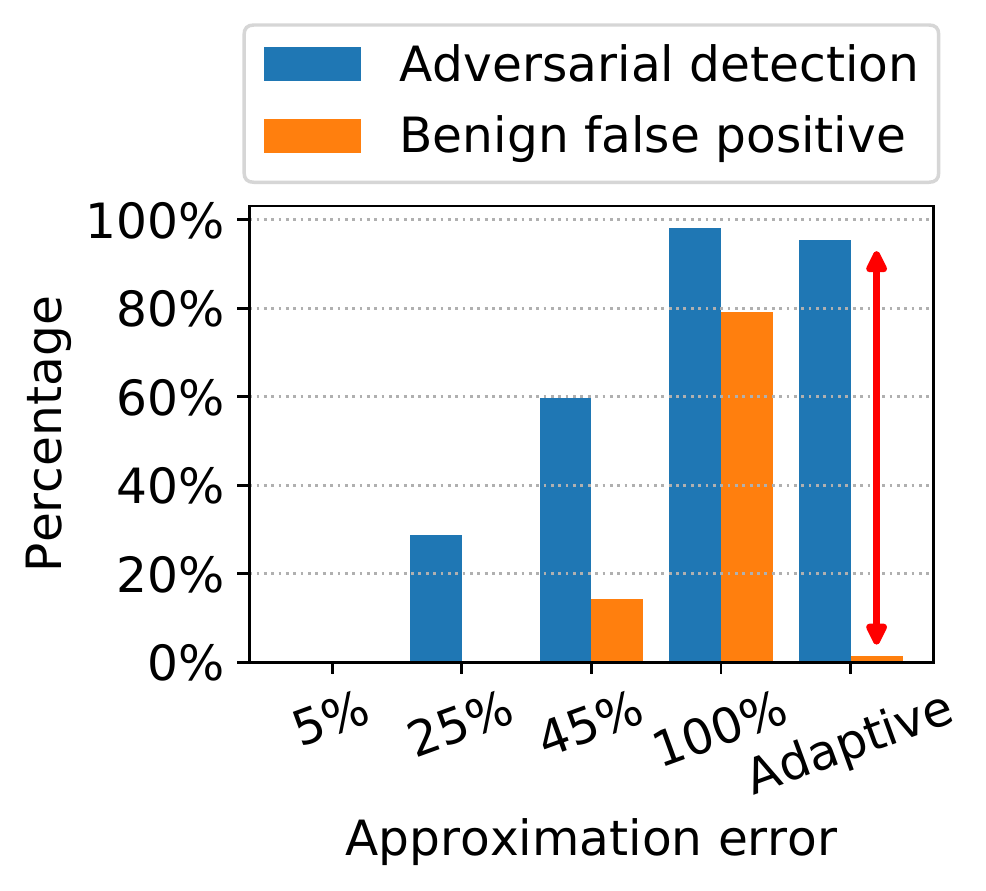}
\caption{Adversarial detection and benign FPR for fixed versus adaptive inference errors in VGG16.}\label{Fig.adaptive_noise}
\end{figure}

The solution we propose in this work is to dynamically adapt the level of error/approximation to the confidence of the classification. To motivate an adaptive approximation over simply increasing the approximation error, we conduct an experiment in which errors are introduced directly into the model, in the activation layers. Figure \ref{Fig.adaptive_noise} shows the adversarial detection rate and benign false positive rate (FPR) for fixed and adaptive error rates. We can see that fixed errors below 50\% have very low detection rates. Increasing the injected error to 100\% results in high detection rate, but at the cost of an unacceptable 80\% false positive rate. The {\em Adaptive} error rate, correlated with the input confidence, achieves both high adversarial detection and low benign FPR. 

The next research question we tackle is how to introduce a well controlled, variable and randomly distributed approximation error into the inference process in a performance-efficient way. Unfortunately, approximate computation using approximate functional units does not offer sufficient flexibility to tune the error rate since they are generally not easily tunable. The same is true for quantization methods, which do not provide sufficient granularity for the approximation errors. In addition, both quantization and approximate computation tends to be deterministic, producing predictable and reproducible error distributions that can be exploited by an attacker.



\section{DNNShield Defense Design}
\label{sec:design}

In order to address the aforementioned challenges we introduce the \tname framework, which includes a new flexible and efficient mechanism to add controlled approximation into the model inference, a method to correlate the amount of error introduced into the model to the confidence of the non-noisy classification and a mechanism for using the approximate inference to detect adversarial inputs. 

\begin{figure}[htbp]
  \centering
     \includegraphics[width=0.8\linewidth]{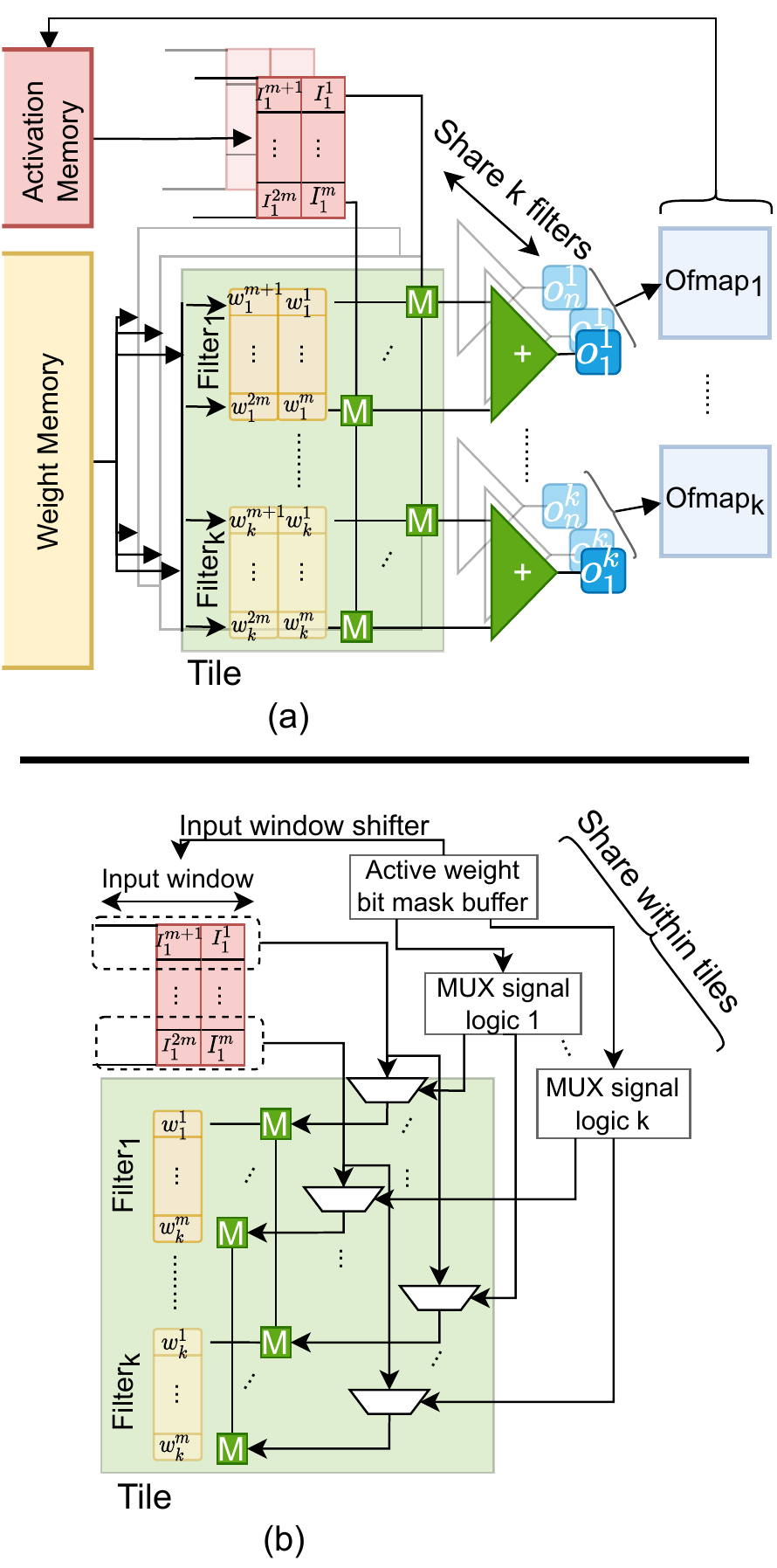}
  \caption{ (a) Baseline accelerator, (b) DNNShield accelerator tile.}\label{Fig.:Dense_Acc}
\end{figure}

\begin{figure*}[htbp]
  \centering
     \includegraphics[width=\linewidth]{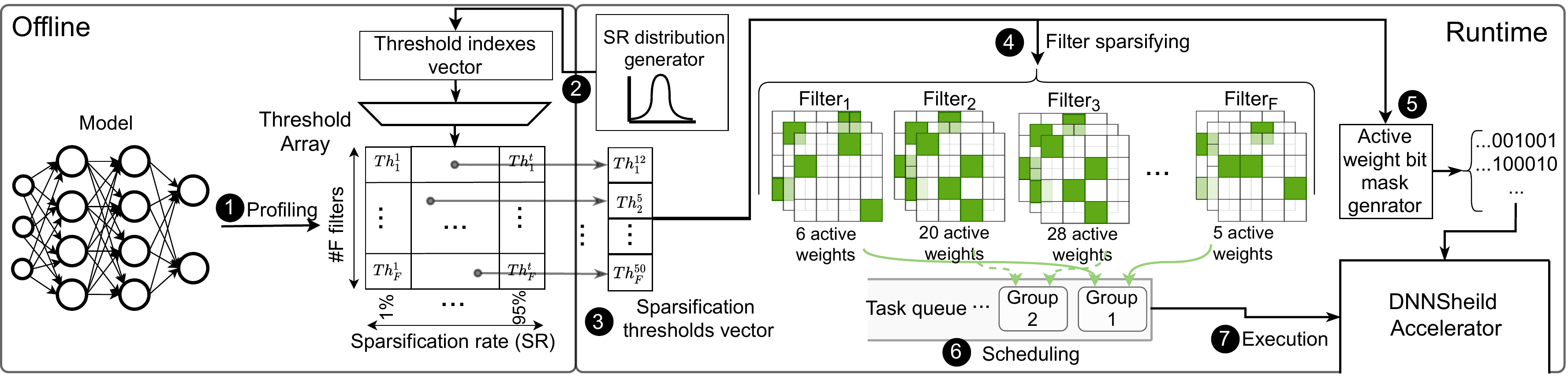}
  \caption{ \tname software consisting of offline model profiling and runtime filter sparsification and scheduling.}\label{Fig.:DySCNN_driver}
\end{figure*}

\subsection{Dynamic Random Sparsification}
\label{noisy_sparsification}

We set a few important criteria for our approximate inference design: (1) flexibility to tune the error rate dynamically at runtime -- to allow error rate to be correlated to input confidence, (2) randomness of the error distribution -- to make defense-aware attacks less likely to succeed, and (3) low overhead. 

In order to satisfy these criteria \tname introduces noise into the DNN by randomly "dropping" (essentially ignoring) weights from the model, through a process we call {\em dynamic random sparsification}. The fraction of dropped weights, or {\em sparsification rate} (SR) controls the amount of error in the model. The sparsification rate for each input is determined based on the classification confidence of the non-noisy run of that input. 


The main advantage of dynamic random sparsification is the potential reduction in performance overhead. Prior work on sparse convolution accelerators \cite{delmas2019bit,zhang2016cambricon,kang2019accelerator,parashar2017scnn,9499811,9499915,9499876,9065523,mahmoud2020tensordash} targeted statically weight sparsed models with the ultimate goal of improving performance/energy efficiency. However, exploiting sparse filters is more challenging in the case of \tname because the filter sparsity changes randomly from run to run. To address these new challenges we developed a hardware/software co-design that profiles the model first and performs scheduling for efficient resource utilization. The hardware accelerator supports dynamic-random sparse execution of the inference with minimum stalls with the help of the software scheduler and flexible control flow.

\subsection{DNNShield Accelerator}
\label{sec:dyscnn}

Figure \ref{Fig.:Dense_Acc} shows the \tname accelerator tile (b) compared to a baseline Dense CNN accelerator (a). The baseline design consists of N tiles which share k filters. Each tile shares the same set of inputs and consists of $k \times m$ MAC units which perform $k \times m$ 8-bit multiplications per cycle. After a total of $k \times M$ ($M=$ filter size) MAC operations, tree-adders accumulate $m$ partial results per output and generate $k$ outputs per tile. The baseline accelerator processes all available weights uniformly, assuming no sparsity. Since the \tname random sparsification is a dynamic process, using conventional sparse accelerators is not practical since they require deterministic offline preprocessing of the statically sparsed model to utilize the resources efficiently at runtime. Our \tname accelerator consists of two components: 1) software scheduler and 2) hardware accelerator. 

\subsubsection{DNNShield Software}
The \tname software handles two main tasks: 1) one-time offline profiling and 2) runtime scheduling (Figure \ref{Fig.:DySCNN_driver}). In order to reduce the overhead of online sparsification, the \tname software parses the model offline and generates a table of threshold values for each filter, corresponding to different sparsification rates \circled{1}. 

\begin{figure}[htbp]
  \centering
     \includegraphics[width=\linewidth]{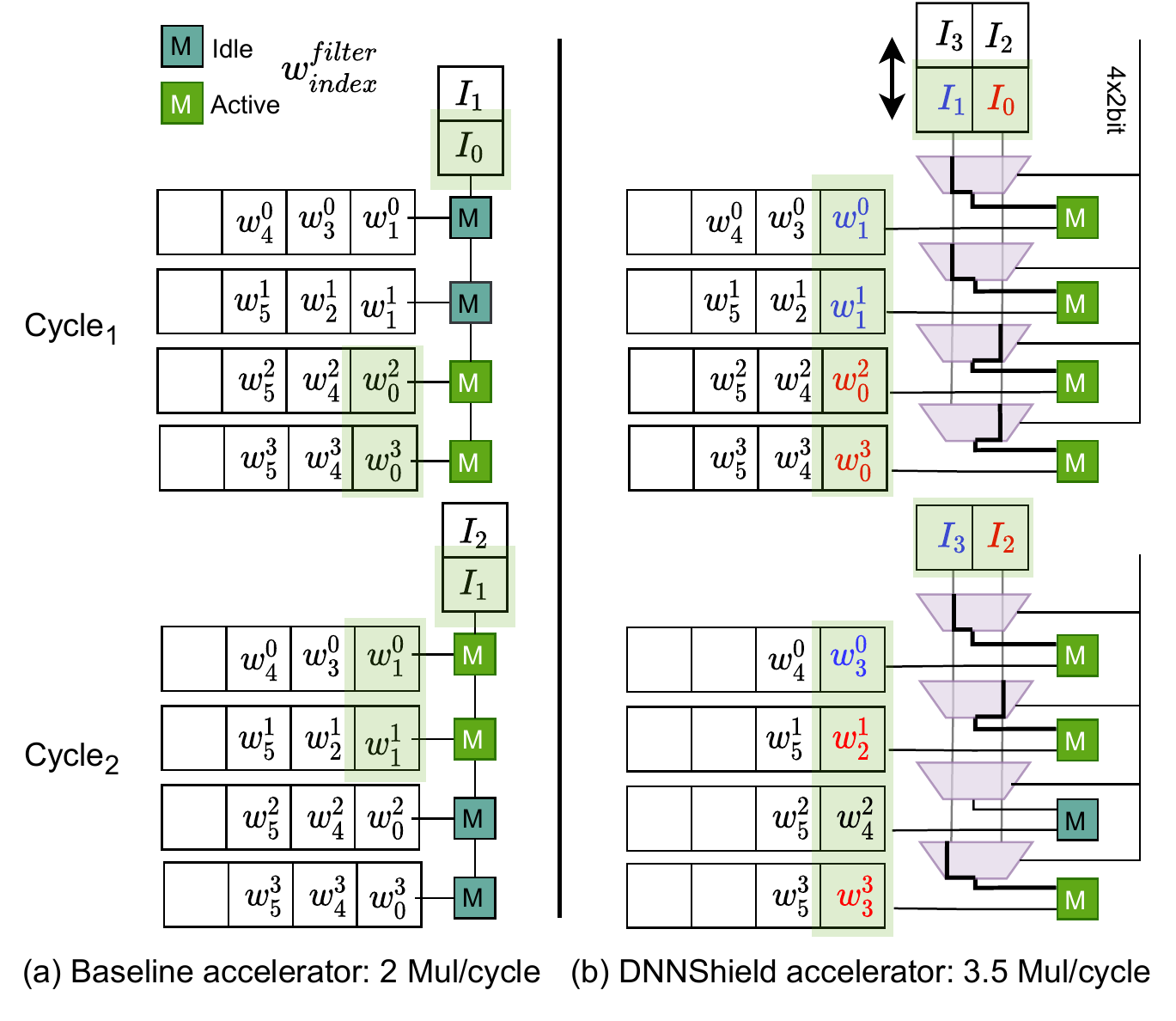}
  \caption{ \tname sparse weights dataflow example through dense baseline (a), \tname accelerator (b).}\label{Fig.:hard_Scheduler}
\end{figure}

At runtime, the {\em SR distribution generator} generates random sparsification rates for each filter \circled{2}, extracts the corresponding threshold value for each filter from the threshold array, and creates a threshold vector \circled{3}. The threshold values are used by the filter sparsifier to drop/ignore weight values below the thresholds assigned to each filter \circled{4}. At the same time, a bit mask of active weights is generated, and will be used by the hardware to map the correct inputs to the active weights \circled{5}. 

Finally, the scheduler groups the filters with roughly similar number of active weights and places them in the task queue \circled{6}. This will increase the efficiency of the inference run since the filters scheduled in the same group will require similar numbers of multiply-accumulate operations, reducing load imbalance and the number of idle cycles. Finally each group is sent to the accelerator together with the active weight bit mask \circled{7}. 



\subsubsection{DNNShield Hardware} The \tname accelerator modifies the baseline dense accelerator to leverage the dynamically sparsified model. The \tname scheduler attempts to schedule the k filters with approximately the same number of active weights. However, \tname needs to make sure that different weights in each kernel can get access to their corresponding input with minimum stalling. For this purpose we used a look-ahead mechanism similar to that in \cite{mahmoud2020tensordash} to match inputs and weights. Figure \ref{Fig.:Dense_Acc}-b shows the \tname accelerator tile. The MUXes are added per each MAC unit to deliver the correct inputs to the active weights in each filter. The accelerator uses the active bit mask to generate the MUX select signals and also identify the offset by which the input window will be shifted every cycle. 


Figure \ref{Fig.:hard_Scheduler} shows an example of how sparse weights and their corresponding inputs flow through for baseline accelerator (Figure \ref{Fig.:hard_Scheduler} (a)) and through \tname (Figure \ref{Fig.:hard_Scheduler} (b)). Figure \ref{Fig.:hard_Scheduler} shows 4 filters sharing the input line within the tile. At cycle 1 $I_0$ is ready to be used by the MAC units sharing the input, however, only two filters have corresponding weights active ($w_0^2$ and $w_0^3$). While the baseline accelerator leaves two MAC units underutilized, \tname fully utilizes the MAC units by performing 4 multiplication at cycle 1 due to the flexibility of selecting appropriate input through MUXes on top of each MAC unit. Therefore $w_1^0$ and $w_1^1$ are consumed by the MAC unit in the same cycle. 
Since no filter needs $I_0$ or $I_1$ the next two inputs $I_2$ and $I_3$ will be loaded into the input buffer in cycle 2. The non-deterministic scheduling of filter groups at runtime prevents pre-generating the signals that drive the selection multiplexers shown in Figure \ref{Fig.:hard_Scheduler} (b). \tname instead uses a "MUX signal logic" unit that uses the bit mask of active weights produced by \tname scheduler to dynamically generate the control signals.

\section{Dynamic Noise and Adversarial Detection}
\label{sec:noise_impact}


\subsection{Impact of Noise Rate on Classification Output}

In order to understand the impact of model approximation on classification output, we use the $L_1$-norm metric to measure the difference between noisy and non-noisy outputs for the same inputs. The $L_1$-norm (also known as Manhattan Distance) is the sum of the absolute pair-wise differences between elements of two vectors. We refer to this difference as the {\em Classification Probability Deviation under Noise} (CPDN). 

\begin{figure}[htpb]
\centering
 \includegraphics[width=0.9\linewidth]{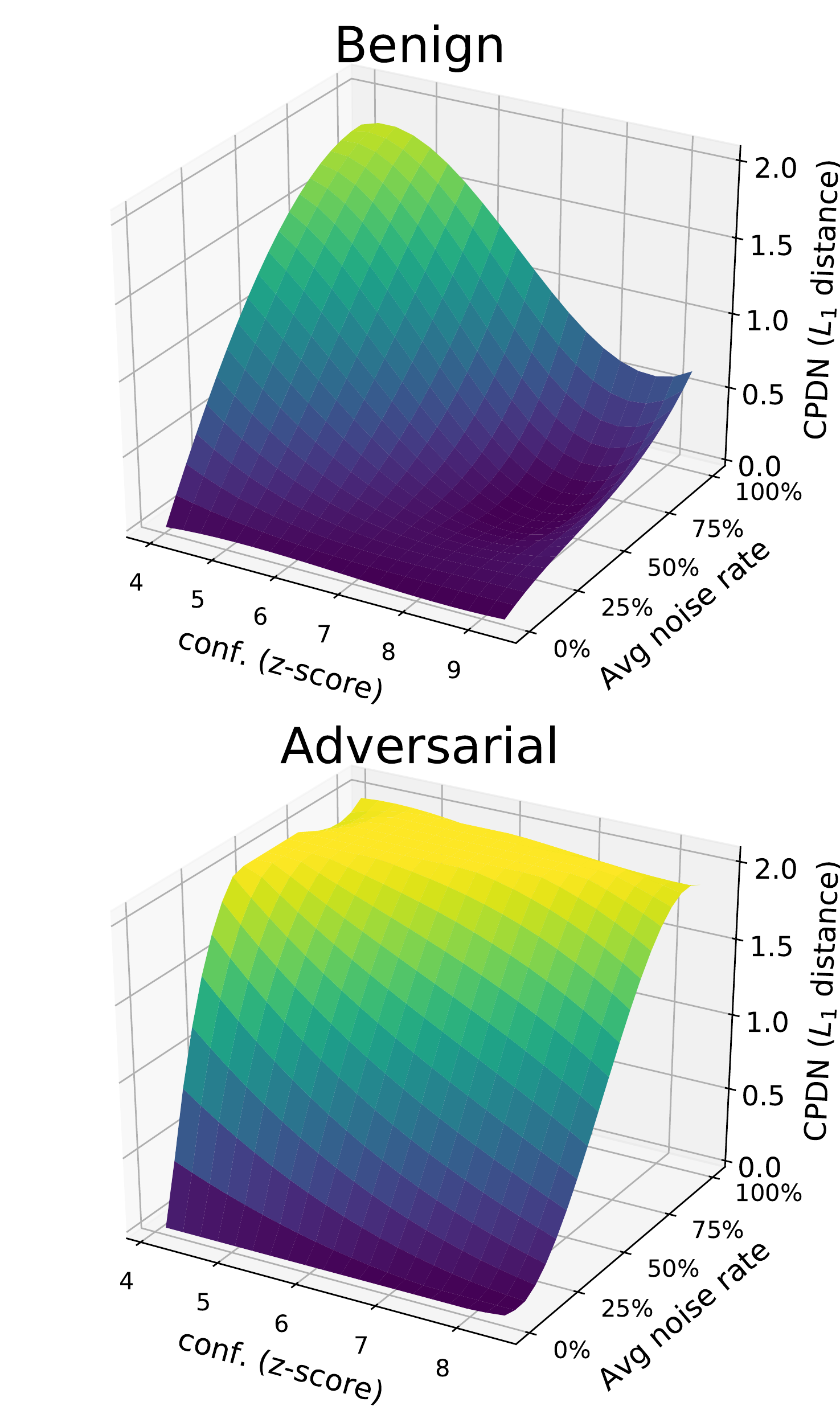}
\caption{Classification Probability Deviation under Noise (CPDN) distribution under variable noise rate and classification confidence for (a) benign and (b) adversarial samples, for VGG16.}\label{Fig.:sensitivity}
\end{figure}


For this analysis we randomly select 1000 benign images from ImageNet. We further generate a total of 1000 adversarial images using 10 different attacks (or attack variants). We run each input 8 times with different noise distributions to generate 8000 sample points for benign as well as adversarial inputs. We measure the CPDN for each input. Figure \ref{Fig.:sensitivity} shows a 3D representation of CPDN for varying classification confidence levels and average noise rates, for benign (a) and adversarial samples (b). Figure \ref{Fig.:sensitivity} shows that across the entire sample space of 8000 data points, CPDN is consistently higher for adversarials compared to benigns, confirming lower robustness to noise for adversarials. 

\begin{figure}[htpb]
\centering
 \includegraphics[width=0.9\linewidth]{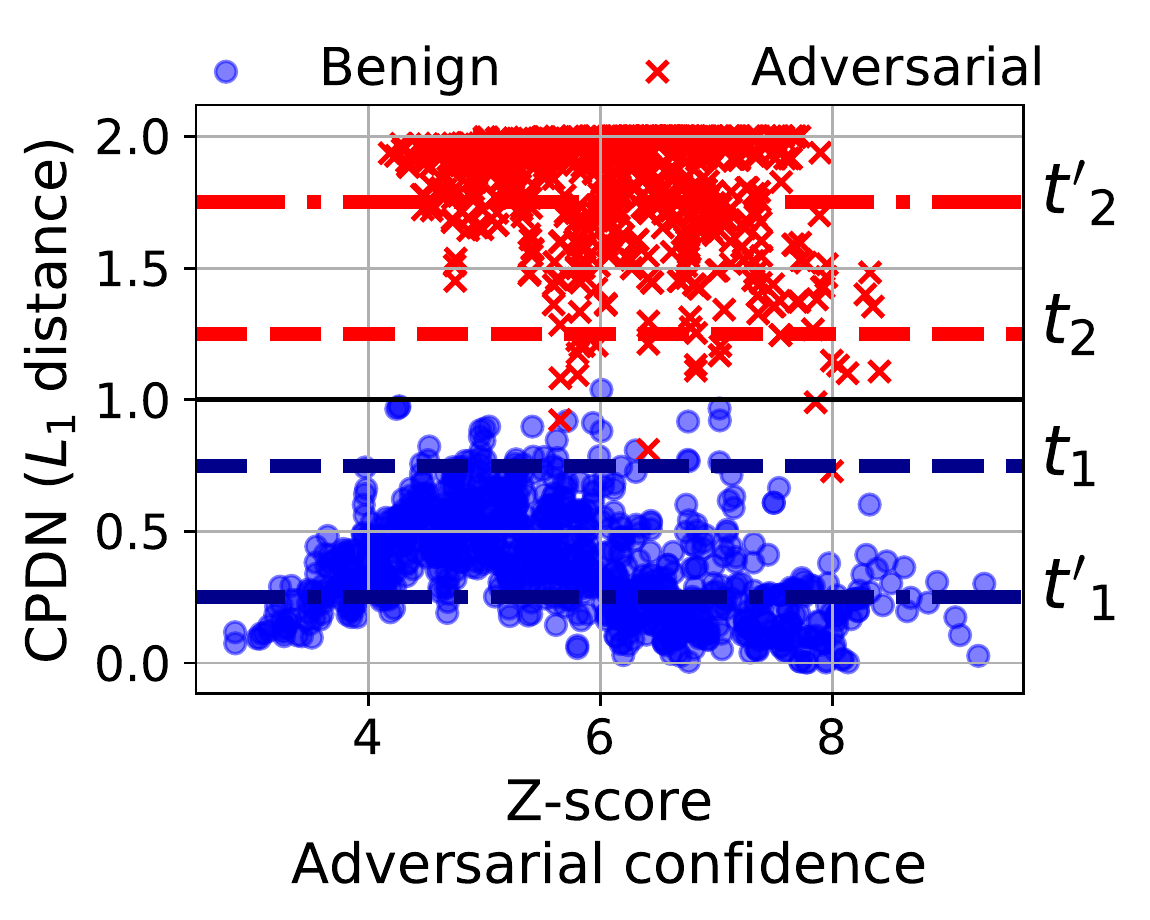}
\caption{ CPDN  distributions  for  benign  and adversarial images when noise is correlated to the confidence of   the   non-noisy classification for VGG16. }\label{Fig.:l1_seperation}
\end{figure}

The data also shows that benign inputs, while more robust to noise than adversarials, are sensitive to high levels of noise if their classification confidence is low. This can be observed in Figure \ref{Fig.:sensitivity} (a) from the high CPDN at high noise and low Z-score. This suggests that low-confidence benign inputs should receive lower noise. Figure \ref{Fig.:sensitivity} (b) shows that low-confidence adversarials exhibit relatively high CPDN even at low noise levels. If we turn our attention to high confidence adversarial samples, we see that they require higher noise levels to exhibit high CPDN. The same high levels of noise, however, when applied to high confidence benigns exhibit a much lower CPDN, indicating that they are robust to high noise. This data suggests that correlating the noise to the confidence of the classification is important to using CPDN in adversarial detection.    

Figure \ref{Fig.:l1_seperation} shows the CPDN distributions for benign and adversarial images when noise is correlated to the confidence of the non-noisy classification -- higher noise for higher confidence. Figure \ref{Fig.:l1_seperation} shows a clear separation between the CPDN distributions of adversarial and benign images, across the entire range of classification confidence that our attacks can generate. As expected, adversarial inputs exhibit higher CPDN deviation from non-noisy baseline compared to benign inputs. \tname uses this separation with appropriate thresholds to detect adversarial inputs.

\subsection{Robustness Analysis}
\label{analysis}

In order to understand why it is important to correlate approximation noise to classification confidence, let us consider a model that classifies input $X$ in the most probable class $C_1$ with probability $P_1$ and the runner-up class $C_2$ with probability $P_2$. Cohen et al.~\cite{cohen2019certified} showed that the distance between $P_1$ and $P_2$ has a direct correlation to the amount of the noise around $X$ that can be tolerated by the classifier. 

The $L_2$ radius $R$ around $X$ can be calculated by:
\begin{equation}\label{eq:robustness}
    R = \frac{\sigma}{2} (\Phi^{-1}(P_1)-\Phi^{-1}(P_2))
\end{equation}
where $\Phi^{-1}$ is the inverse of standard Gaussian CDF and $\sigma$ is the standard deviation of the noise. The higher the $R$ value for an input $X$, the more noise the classifier can tolerate and still classify $X$ correctly. According to Equation \ref{eq:robustness} the radius $R$ is large when the probability of top class $C_1$ is high and the probability of the next class is low, which corresponds to high confidence classification. This shows robustness to noise is correlated to classification confidence. We therefore expect that benign images will not suffer from high false positive rate regardless of confidence, if the approximation noise is scaled with the confidence.

\subsection{Adversarial Detection}

\begin{figure}[htb]
\centering
 \includegraphics[width=\linewidth]{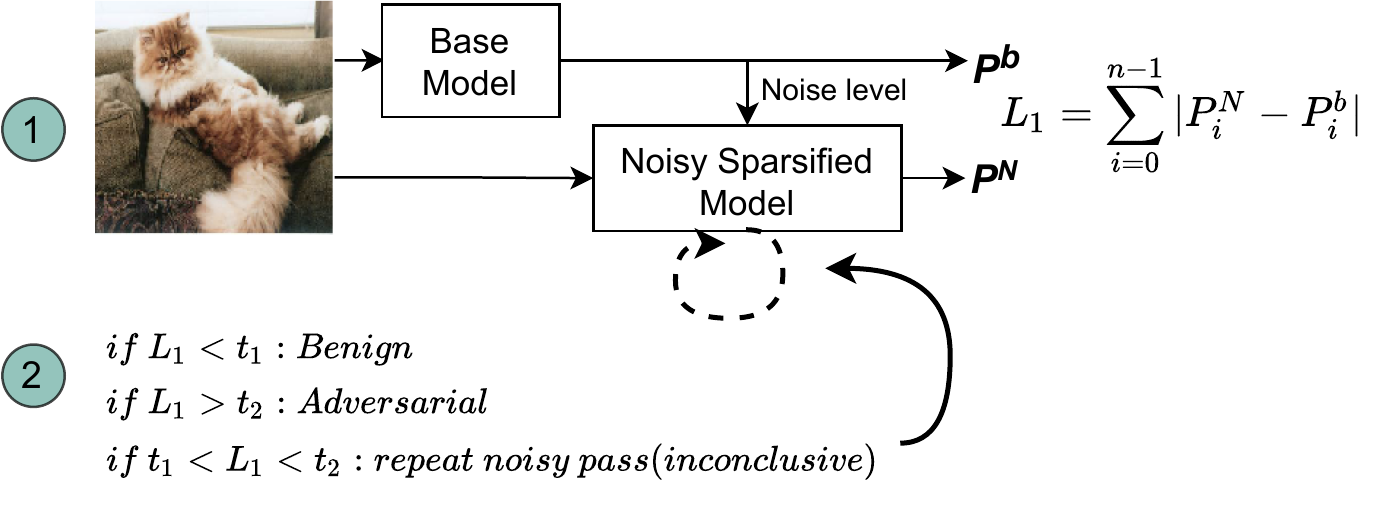}
\caption{ Adversarial detection using CPDN ($L_1$ distance).}\label{Fig.:detection}
\end{figure}

The \tname framework relies on the CPDN metric to detect adversarial inputs. This process is depicted in Figure \ref{Fig.:detection}. Initially, a first inference pass through the network is performed without noise injection to establish a reference. The output classification is recorded as $P^b$. This is followed by another approximate inference pass. The confidence of the classification $P^b$ is used to determine the amount of noise to be injected.  

The $L_1$ distance between the output vectors of the noisy ($P^N$) and non-noisy ($P^b$) inference passes is computed. The $L_1$ distance is then compared with different thresholds values.  Depending on the outcome, subsequent inference passes may be required. Figure \ref{Fig.:l1_seperation} shows the 4 thresholds used by the detection mechanism overlaid on the $L_1$ distance distribution for \textit{VGG16}. 
$t_1$ and $t_2$ represent the $L_1$ distance below/above which most benign/adversarial images fall, respectively. $t'_1$ and $t'_2$ represent tighter thresholds below/above which about 80\% of the benign/adversarial images fall. 

If the measured $L_1$ distance is either very high or very low ($<t'_1$ or $>t'_2$) the input image can immediately, and with high confidence, be classified as benign or adversarial, respectively. Most inputs  ($>\approx80\%$) from both our benign and adversarial test sets fall in this category. In this case the detection algorithm terminates and the outcome is reported.
 
Otherwise, \tname cannot yet make a high-confidence detection, and another noisy inference pass is required. The average $L_1$ distance over all the previous noisy runs is computed and compared with the more conservative thresholds $t_1$ and $t_2$. The images with average $L_1 < t_1$ are classified as benign and those with average $L_1 > t_2$ are classified as adversarial. If $t_1 < L_1 < t_2$, a new noisy pass is performed and the average $L_1$ distance is recomputed. The algorithm repeats until a maximum $M$ number of iterations is reached ($M=4$ and $M=8$ in our experiments). If a classification is still not possible, the algorithm defaults to Benign. 

\section{DNNShield Prototype Implementation}
\label{sec:prototype}

\begin{figure}[htbp]
\centering
     \includegraphics[width=0.9\linewidth]{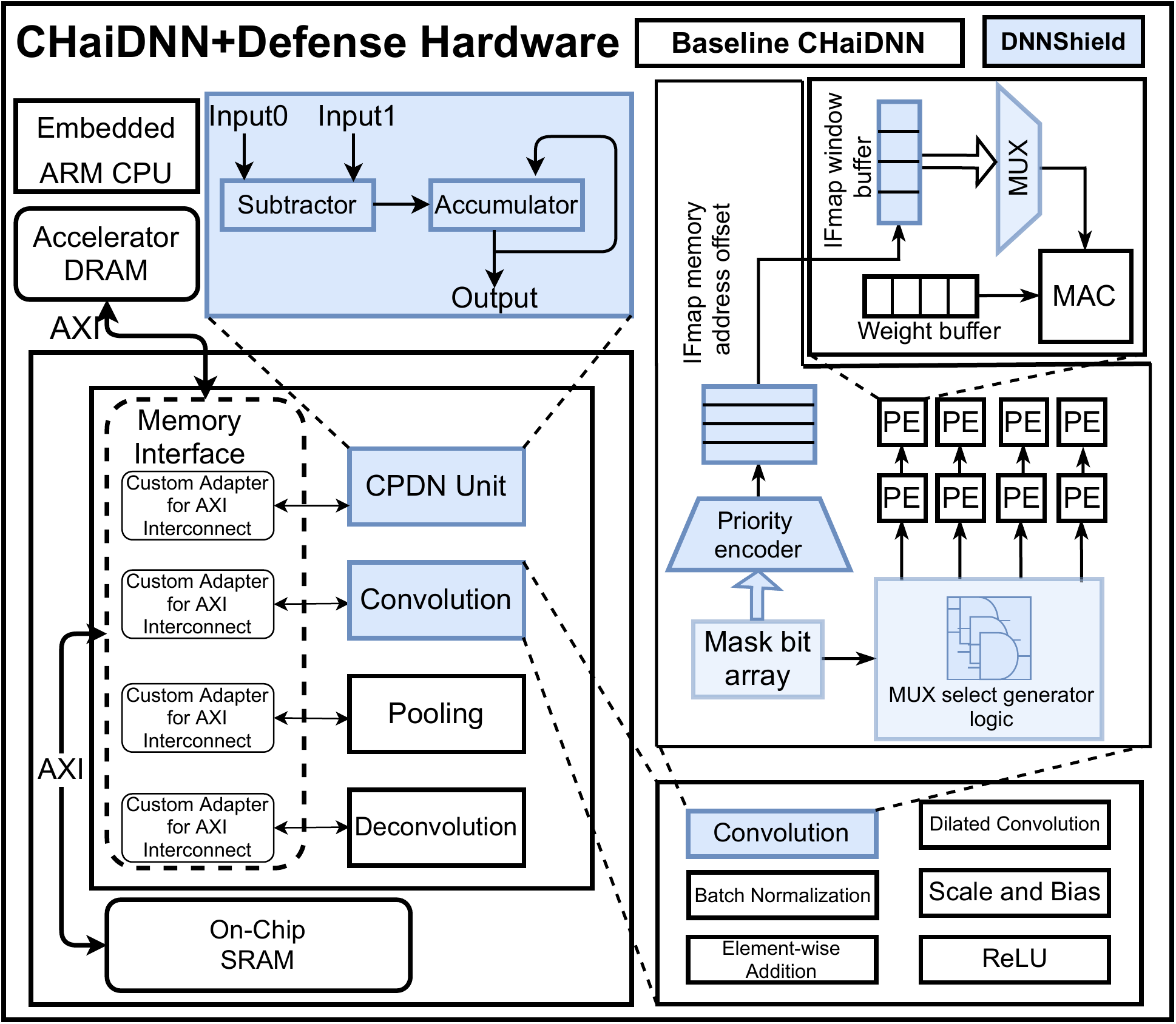}
  \caption{\tname hardware design based on the Xilinx CHaiDNN accelerator.}\label{fig:hasi_HW}
\end{figure}

As a proof-of-concept, we implement \tname in a FPGA-based DNN accelerator, the Xilinx CHaiDNN architecture \cite{chaidnn}. Figure \ref{fig:hasi_HW} shows a diagram of the design. The baseline includes dedicated hardware for Convolution Pooling, and Deconvolution functions. All the compute elements are connected to a Memory Interface Block which allows access to the on-chip SRAM as well as the main system DRAM via a custom AXI Interconnect. \tname augments the baseline accelerator with the following components color coded blue: (1) modified convolution supporting noisy sparsification including MUX select generator logic, input window buffers and priority encoders, (2) custom logic for computing CPDN, and (3) additional control logic for coordinating partial result reuse and early termination.  

\begin{figure}[htbp]
\centering
     \includegraphics[width=0.9\linewidth]{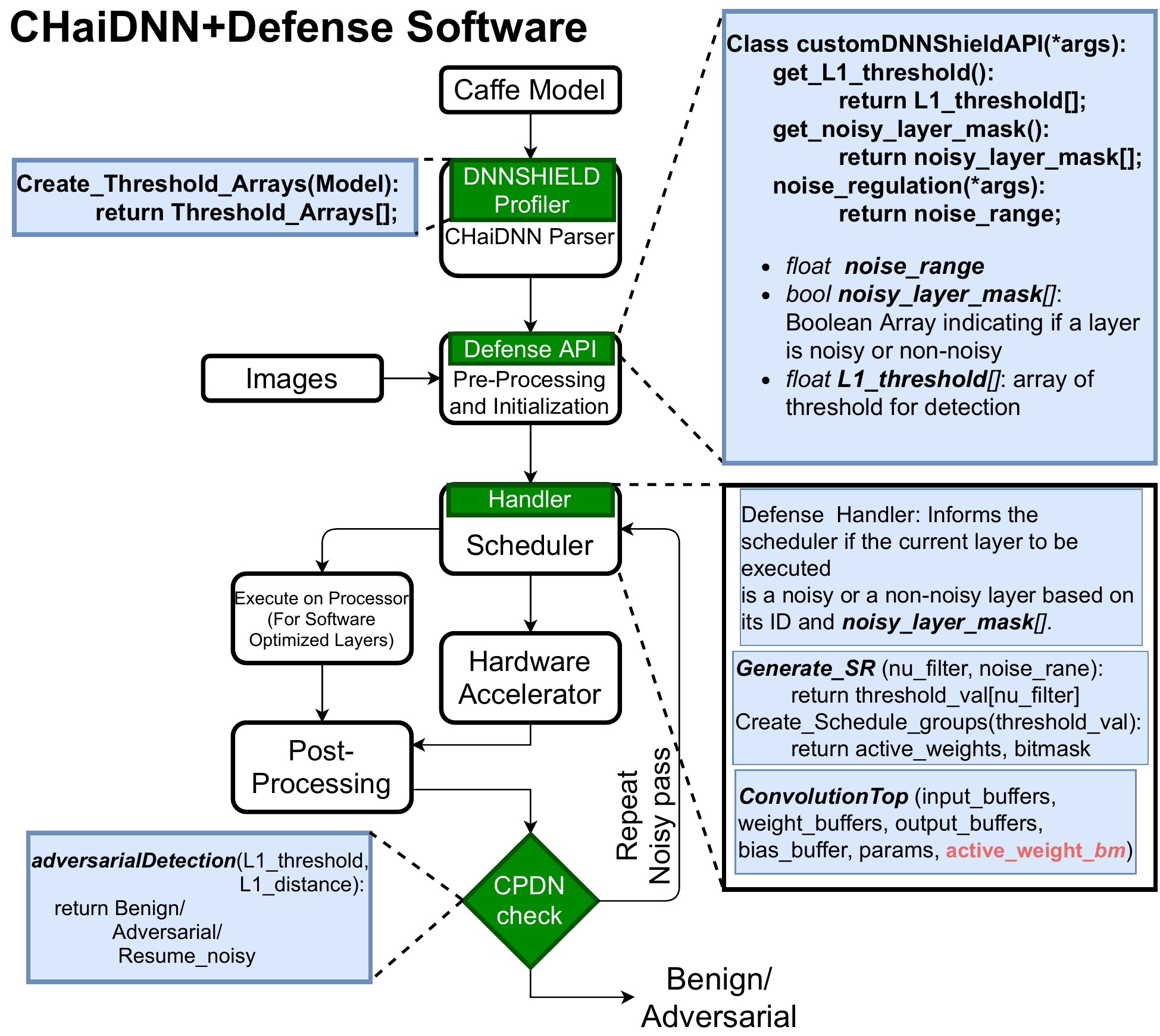}
  \caption{ \tname Runtime integrated with CHaiDNN software.}\label{fig:hasi_SW}
\end{figure}

\noindent{\bf Convolution} is the core of noisy sparsification with MUXes distributed through PEs for and MUX signal generators. PEs within the column share MUX signals. Also, priority encoders are used to determine the number of inputs that need to be bypassed regarding the ignored weights in sparsification. Then the address offset is calculated and the next set of inputs will be loaded in the input buffer.   

\noindent{\bf CPDN Unit} 
is used to compute the CPDN between a noisy and non-noisy output. It primarily consists of logic to subtract vector elements and accumulate the absolute value of the result, shown in Figure \ref{fig:hasi_HW}. 

\noindent{\bf DNNShield Runtime:} The CHaiDNN/\tname software stack as shown in Figure \ref{fig:hasi_SW}, includes of a parser for the input model and a pre-processor for the DNN inputs. \tname profiler is added to the ChaiDNN parser to create the threshold array that later will be used by \tname. 
The \tname Runtime is invoked by the pre-processor following the initial non-noisy run to determine the noise level to be injected. 
The \tname handler is designed to invoke the scheduler when the CPDN needs to be computed in the hardware. 
It also augments the scheduler with \tname scheduler for grouping the balanced filter together for more efficient execution. 


\section{Methodology} 
\label{sec:method}


We implement, synthesize and deploy \tname on CHaiDNN running on a Xilinx Zynq UltraScale+ FPGA. The SoC associated with the board is ZU7EV which integrates a quad-core Arm Cortex-A53 processor. CHaiDNN is an open-source Deep Neural Network accelerator designed to run on Xilinx Ultrascale MPSoCs. 
We compare our FPGA accelerator with two software implementation of \tname on a CPU and GPU. We implemented the software \tname using TensorFlow2 \cite{tensorflow}. We run our software \tname on Intel Core-i7 CPU@3.40GHz and NVIDIA RTX-2060 Turing GPU.

\subsection{DNN Models, Input Dataset, Attacks}

We used two networks VGG16 \cite{simonyan2014deep} and ResNet50 \cite{ResNet} trained on ImageNet \cite{krizhevsky2012imagenet} for running attacks and generating adversarial images. 
Targeted attacks, which aim to misclassify an input into a target class, use two types of targets called \textit{Next} and \textit{LL}. \textit{Next} corresponds to choosing the target class $t=L+1~mod~\#classes$  where L is the sorted index of top ground truth classes. For \textit{LL} the target class $t$ is chosen as least likely class ($t=min(\hat{y})$) where $\hat{y}$ is the prediction vector of an input image. 
Table \ref{tab:attack_stats} summarizes the adversarial attacks alongside their detailed parameters, success rate, average confidence and average distortion with different metrics per model. 

\subsection{Comparison with Existing Defenses}

We compared our detection rate of adversarial as well as True/False Positive rate (TPR/FPR) of benign images with two state-of-the-art post-training defense mechanisms, detailed below. 
\textbf{Stochastic Activation Pruning (SAP)} \cite{dhillon2018stochastic} introduces
randomness into the evaluation of a neural network to defend against adversarial examples. 
SAP randomly drops some neurons of each layer
to 0 with a probability proportional to their absolute value. 
Values which are retained are scaled up to retain accuracy. 
\textbf{Feature squeezing (FS)} \cite{xu2018feature,liang2017detecting} is a correction-detection mechanism that relies on reducing the input space (and attack surface) by "squeezing" images.
FS requires off-line profiling and training to find the best squeezer and corresponding thresholds for each pair of data-set and attack, making it less practical to deploy in real-world applications. FS also requires at least 3 squeezers, resulting in at least 4$\times$ performance overhead. 
For a fair comparison we retrained FS on our set of benign and Adversarial images for both VGG16 and ResNet50 separately. 
\textbf{Approximate mul8u\_KEM:} We  also  compared with an approximate multiplier ({\em Approx. mul8u\_KEM}) from \cite{evoapprox16}, in an approach similar to \cite{guesmi2020defensive}.

\begin{table*}
\caption{Attack parameters for multiple variants of CW and EAD attacks. Original attack success rate, confidence, and distortion. Detection rates for 4 defenses: SAP, FS, Approximate-MUL and \tname. Dataset from ImageNet.}
\label{tab:attack_stats}
\small
\centering
\scalebox{0.6}{
\begin{tabular}{|l|c|ad|ad|ad|ad|ad|ad||ad|ad|ad|ad|ad|}
\hline
\multicolumn{14}{|c||}{\textbf{Attacks}} & \multicolumn{8}{c|}{\textbf{Defenses}}\\\hline\hline  
\multirow{3}{*}{ Attack} & \multirow{3}{*}{ Target} &\multicolumn{2}{c|}{\multirow{2}{*}{ Param k} } &  \multicolumn{2}{c|}{\multirow{2}{*}{ Mean Confidence}} & \multicolumn{8}{c||}{ Distortion}  & \multicolumn{2}{c|}{\multirow{2}{*}{ SAP$^*$}} & \multicolumn{2}{c|}{\multirow{2}{*}{ FS$^+$}} & \multicolumn{2}{c|}{Approximate} & \multicolumn{2}{c|}{\multirow{2}{*}{\tname}} \\
\cline{7-14}
       &&  \multicolumn{2}{c|}{} & \multicolumn{2}{c|}{} & \multicolumn{2}{c|}{\large $L_0$}  & \multicolumn{2}{c|}{\large $L_1$} & \multicolumn{2}{c|}{\large $L_2$} & \multicolumn{2}{c||}{\large $L_{\infty}$} &   \multicolumn{2}{c|}{} &  \multicolumn{2}{c|}{} & \multicolumn{2}{c|}{mul8u\_KEM} &\multicolumn{2}{c|}{}  \\
       \cline{3-22}
       && VGG & RNet  & VGG & RNet & VGG & RNet & VGG & RNet & VGG & RNet & VGG & RNet & VGG & RNet & VGG & RNet & VGG & RNet & VGG & RNet\\\hline
\multirow{2}{*}{\textbf{$CW_{L0}$} } &    Next     & 5& 5 &                                92.9\% &                    94.5\% &       42.2\% &          42.0\% &         232 &            105 &       10.44 &           6.26 &               0.94 &                  0.88 &   35\% &      42\% &  \textbf{\underline {100\% }}&     \textbf{\underline {100\%}} &    34\% &      59\% &       67\% &      82\% \\
           & LL &    &      &                               84.8\% &                    87.9\% &       42.4\% &          42.1\% &         382 &            185 &       13.65 &           8.33 &               0.96 &                  0.92 &   29\% &      43\% &  100\% &     100\% &    56\% &      96\% &      \textbf{\underline {100\%}} &     \textbf{\underline {100\% }}\\
           
\hline
\multirow{4}{*}{\textbf{$CW_{L2}$}}&  Next & 10 & 30 &                               100.0\% &                   100.0\% &      100.0\% &         100.0\% &         412 &            411 &        1.69 &           1.63 &               0.07 &                  0.06 &   45\% &      45\% &   84\% &      89\% &    68\% &      11\% &      \textbf{\underline { 91\% }}&     \textbf{\underline {100\% }}\\


            &  LL &   &      &                                  100.0\% &                   100.0\% &      100.0\% &         100.0\% &         555 &            531 &        2.24 &           2.07 &               0.08 &                  0.06 &   36\% &      46\% &  100\% &    100\% &    81\% &      28\% &      \textbf{\underline {100\% }}&     \textbf{\underline {100\% }}\\
\cdashline{2-22}
          & Next &70&140&                              100.0\% &                   100.0\% &       99.9\% &         100.0\% &       2,015 &          2,609 &        7.43 &           9.22 &               0.24 &                   0.2 &    0\% &      21\% &    6\% &      48\% &     7\% &       0\% &       \textbf{\underline {84\% }}&      \textbf{\underline {89\% }}\\
            &LL  &    &       &                                 100.0\% &                   100.0\% &       99.9\% &         100.0\% &       2,176 &          3,267 &        8.07 &          11.56 &               0.25 &                  0.24 &    0\% &      17\% &    9\% &      67\% &    19\% &       0\% &       \textbf{\underline {96\% }}&      \textbf{\underline {98\% }}\\
            
\hline
\multirow{2}{*}{\textbf{$CW_{L\infty}$}}   &  Next & 5 & 5&                             94.7\% &                    95.3\% &      100.0\% &         100.0\% &         735 &            523 &        2.27 &            1.6 &               0.01 &                  0.01 &   42\% &      47\% &   \textbf{\underline {91\% }}&      \textbf{\underline {96\% }}&    69\% &      71\% &       83\% &      89\% \\
     & LL &    &       &                                         91.8\% &                    93.4\% &       99.9\% &         100.0\% &       1,023 &            699 &        3.05 &           2.12 &               0.01 &                  0.01 &   45\% &      45\% &  100\% &     100\% &    94\% &      98\% &      \textbf{\underline {100\% }}&     \textbf{\underline {100\% }}\\

\hline
\multirow{4}{*}{\textbf{$EAD_{L1}$}} &Next &10&30&                              100.0\% &                   100.0\% &       52.4\% &          58.3\% &         173 &            205 &        2.69 &           2.79 &               0.24 &                  0.22 &   27\% &      52\% &   78\% &      98\%&    41\% &       2\% &       \textbf{\underline {91\% }}&      \textbf{\underline {98\% }}\\

   & LL  &   &       &                                 100.0\% &                   100.0\% &       54.7\% &          59.4\% &         269 &            276 &        3.56 &           3.47 &               0.29 &                  0.26 &   34\% &      41\% &  100\% &      98\% &    59\% &       0\% &      \textbf{\underline {100\% }}&     \textbf{\underline {100\% }}\\

\cdashline{2-22}
 & Next & 70& 140&                                   100.0\% &                   100.0\% &       78.8\% &          80.5\% &       1,400 &          1,734 &        9.92 &          12.34 &               0.54 &                  0.49 &    0\% &      16\% &    4\% &      46\% &     2\% &       0\% &       \textbf{\underline {78\% }}&      \textbf{\underline {88\% }}\\
 
  & LL  &   &      &                                       100.0\% &                   100.0\% &       78.3\% &          85.5\% &       1,510 &          2,782 &       10.65 &          17.36 &               0.55 &                  0.58 &    0\% &      18\% &    4\% &      81\% &     4\% &       0\% &      \textbf{\underline { 93\%}} &     \textbf{\underline { 84\% }}\\
  
\hline
\multirow{2}{*}{\textbf{$EAD_{EN}$}}& Next & 10& 30 &                                100.0\% &                   100.0\% &       47.9\% &          52.4\% &         191 &            421 &        4.36 &           6.73 &                0.6 &                  0.73 &   17\% &      40\% &   63\% &      89\% &    27\% &       0\% &      \textbf{\underline { 80\% }}&      \textbf{\underline {94\% }}\\
    & LL  &   &       &                                  100.0\% &                   100.0\% &       48.0\% &          57.4\% &         252 &            596 &         5.9 &           8.59 &               0.72 &                   0.8 &   12\% &      26\% &   \textbf{\underline {98\%}} &      98\% &    36\% &       2\% &       \textbf{\underline {99\% }}&      97\% \\

\hline
\multicolumn{4}{|l}{RNet: ResNet50, VGG: VGG16}&\multicolumn{9}{l}{$^+$FS threshold: VGG16: 1.022, ResNet50: 1.229,}&\multicolumn{1}{r||}{AVG}&   16.6\% &      33.2\% &    55\% &       79\% &    36\% &       15\% &  \textbf{\underline {86\% }}& \textbf{\underline {88\%}}\\
\cline{16-24}
\multicolumn{5}{|r}{$^*$Values for SAP are accuracy~~~~~~~~~~~}&\multicolumn{9}{r||}{FPR}&   58\% &      37\% &    6\% &       3\% &    29\% &       19\% & 6\% & 6\%\\
\hline

\end{tabular}
}
\end{table*}
\section{Evaluation}
\label{sec:eval}

We evaluate \tname adversarial detection rate, robustness to defense-aware attacks, and performance and area overheads. We also conduct a number of sensitivity studies for the main design parameters. 

\subsection{Adversarial Detection} 
We first look at \tname's ability to identify adversarial images. We measure the detection rate for adversarial inputs as well as the false positive rate (FPR) for benign inputs. We compare \tname with Feature Squeezing (FS) and SAP for multiple configurations of CW and EAD. Table \ref{tab:attack_stats} lists the detection rate for all the attack variants we evaluate, for both VGG and ResNet. 

The results show that both \tname significantly outperforms both FS and SAP on average. \tname shows an average detection rate of 86\% and 88\% for VGG and ResNet, respectively. \tname also significantly outperforms the state of the art defense, FS which averages 55\% and 79\% for VGG and ResNet, respectively. This is especially true for high-confidence attack variants, for which FS does not work as well. For instance, under the $EAD_{L1}$ attack with $k=70$ we see 93\% detection rate for \tname vs. 4\% for FS (VGG16). This shows that \tname is resilient to very strong attacks. 



\begin{figure}[htbp]
  \centering
    \includegraphics[width=0.9\linewidth]{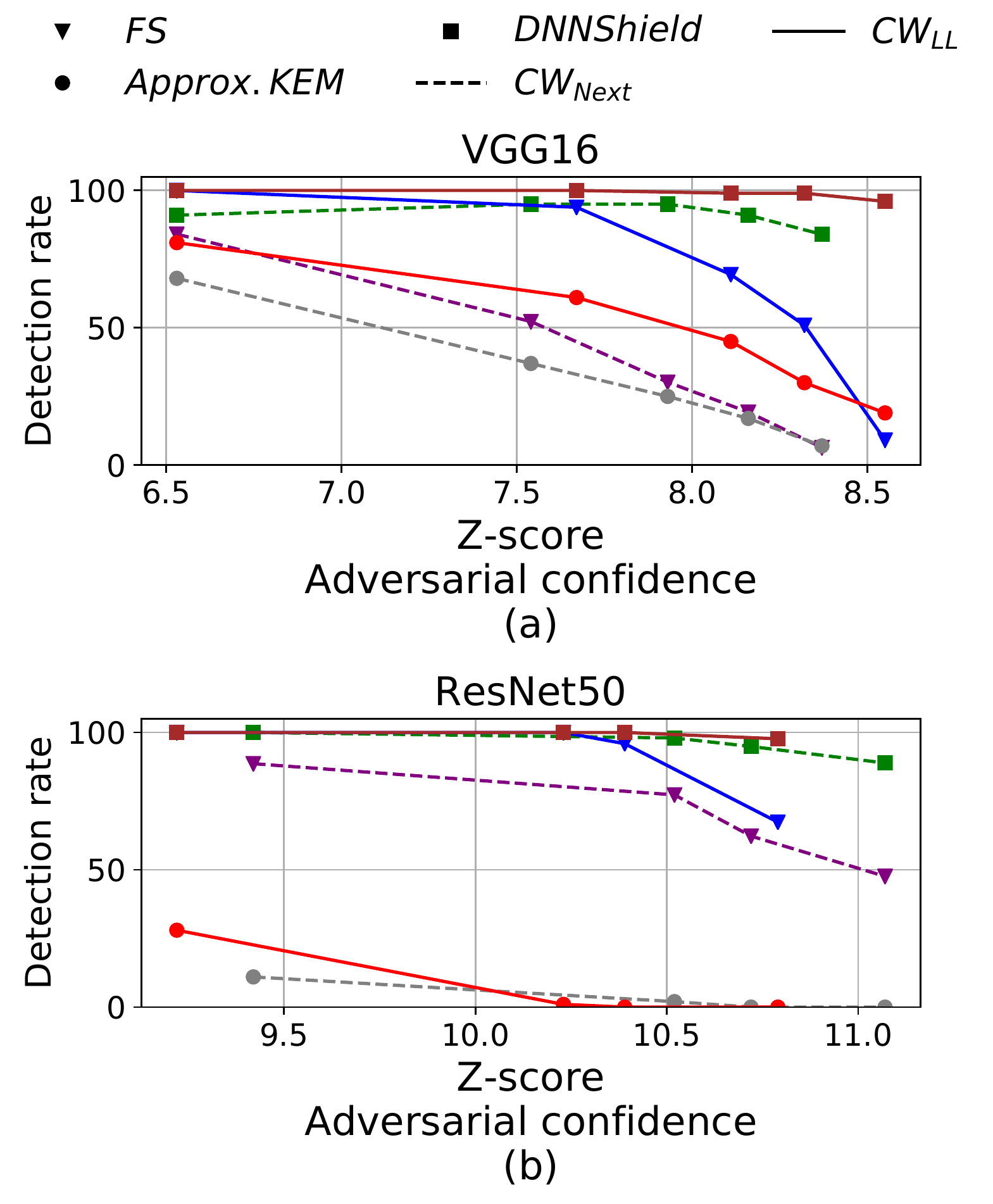}
  \caption{ Detection rate for different adversarial confidence generated by $CW_{L2}$ attack, (a) VGG16 and (b) ResNet50}\label{Fig.:strong_attack}. 
\end{figure}

Figure \ref{Fig.:strong_attack} shows detection rate versus adversarial confidence for \tname, FS and the {\em Approx. mul8u\_KEM} as a function of classification confidence. Both FS and {\em Approx. mul8u\_KEM} detection rates fall steeply as confidence increases while \tname detection rate remains high. These results re-emphasize the importance of adapting the approximation error to the confidence of the classification.


\subsection{DNNShield-Aware Attacks}
\label{sec:hasi-aware}

In order to investigate the robustness of the \tname defense, we construct a set of attacks tailored specifically to defeat it. These attacks assume full knowledge of the \tname design. In theory, \tname could be defeated by an attack that generates adversarial examples for which the model's robustness to approximate inference is similar to that of benign examples. In order to attempt to generate such adversarial examples, we used the approach suggested in \cite{tramer2020adaptive} to generate adversarial examples that target the probability vector of an arbitrary benign example from another class. The idea is to create an adversarial example that mimics the response of benign images under noise. Hence, for sample $x$ of class $y$, we pick a target $t\neq y$ and create adversarial example $x'$ that minimizes the objective:
\begin{equation} 
    \textrm{$minimize$}~~~||y(x')-y(x_t)||_1
\end{equation}
where $y(x')$ and $y(x_t)$ are the probability vector of the adversarial and target inputs respectively. While we try to minimize the $L_1$ distance between adversarial and the benign target, we need to also minimize the adversarial perturbation under the $L_2$  distortion metric. The final objective function is:
\begin{equation}\label{eq:minimize}
\begin{split}
\textrm{$minimize_x$}~~\;cf(x,t) +\beta||y(x')-y(x_t)||_1+ ||x-x_0||_2 ^2\\
\textrm{such that }\;x  \in [0,1]^n\\
\end{split}
\end{equation}
where $f(x,t)$ denotes the loss function and $\beta$ is the regularization parameter for $L_1$ penalty. Increasing $\beta$ forces a lower $L_1$ distance between the adversarial and target benign and could evade \tname detection.

\begin{figure}[htbp]
\centering
 \includegraphics[width=0.9\linewidth]{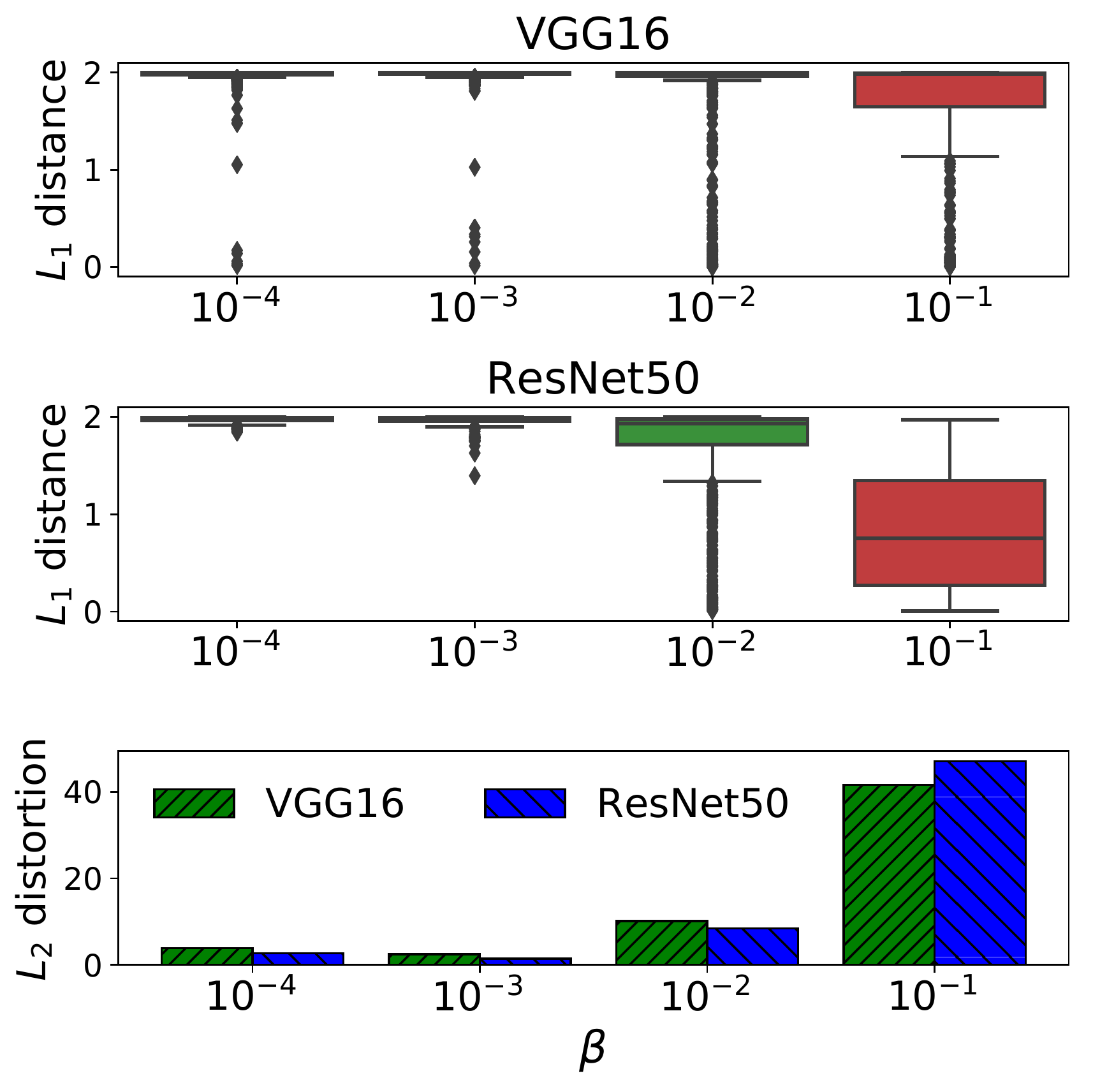}
\caption{ $L_1$ distance vs. $L_2$ distortion for different $\beta$ values.}\label{Fig.:beta}
\end{figure}


Table \ref{tab:adaptive} summarizes the adaptive attack parameters and detection rates under \tname. We can see that for low-$\beta$ attacks, \tname detection rate is very high ($A_{1}$-$A_{3}$). For $A_{4}$, with $\beta=10^{-1}$ the detection rate is lower. However, in order to generate $A_{4}$, the $L_2$ distortion has to be increased by 4-5$\times$ relative to $A_{3}$. To understand why, Figure \ref{Fig.:beta} shows the effect of $\beta$ on the $L_1$ distance of probability distribution and $L_2$ distortion. Optimizing for both low $L_2$ distortion and $L_1$ distance are competing objectives. Increasing $\beta$ will decrease the $L_1$ distance, making the adversarial harder to detect, but it also increases $L_2$ distortion. 
The target benign input, which the adversarial sample is trying to mimic, is chosen randomly from 1000 images in the adversarial targeted class. While a few of these targets do lead to lower distortion, the average distortion, for high $\beta$ ($10^{-1}$), is very high. Another popular approach is using an EoT attack in which noise (transformation) was applied during adversarial generation. We injected variable noise correlated to the confidence of the classification in each training iteration, as in DNNShield. The result was that, because of the variable noise, the attack could not converge on a successful adversarial. Using fixed noise as in traditional EoT did not work either because of the adaptive DNNShield response.

\begin{table}[htbp]
\centering
\caption{\tname aware adaptive attacks} 
\label{tab:adaptive}
\scalebox{0.75}{
\begin{tabular}{|c|c|ad|ad|ad|ad|}
\hline

 & \multirow{2}{*}{$\beta$} & \multicolumn{2}{|c}{Success rate} & \multicolumn{2}{|c}{Mean Confidence} & \multicolumn{2}{|c}{$L_2$ Distortion}   & \multicolumn{2}{|c|}{\tname det.}  \\
\cline{3-10}
Attack &             &    VGG  &    RNet         &     VGG                 &    RNet                   &  VGG   &       RNet      &       VGG        &    RNet             \\
\hline
 $A_1$  & $10^{-4}$    &               100\% &                  100\% &                 94.3\% &                    95.9\% &        3.91 &           2.71 &   99\% &     100\% \\
   
$A_2$ &$10^{-3}$      &               100\% &                  100\% &                 92.3\% &                    94.0\% &        2.48 &           1.42 &   99\% &     100\% \\
 
$A_3$ &$10^{-2}$      &               100\% &                  100\% &                 96.1\% &                    96.9\% &       10.14 &           8.45 &   95\% &      84\% \\
 
 $A_4$  &    $10^{-1}$ &        58\% &                   31\% &                 97.7\% &                    97.9\% &       41.58 &          46.99 &   81\% &      39\% \\
\hline
\end{tabular}
}
\end{table}

\begin{figure}[htbp]
\centering
 \includegraphics[width=\linewidth]{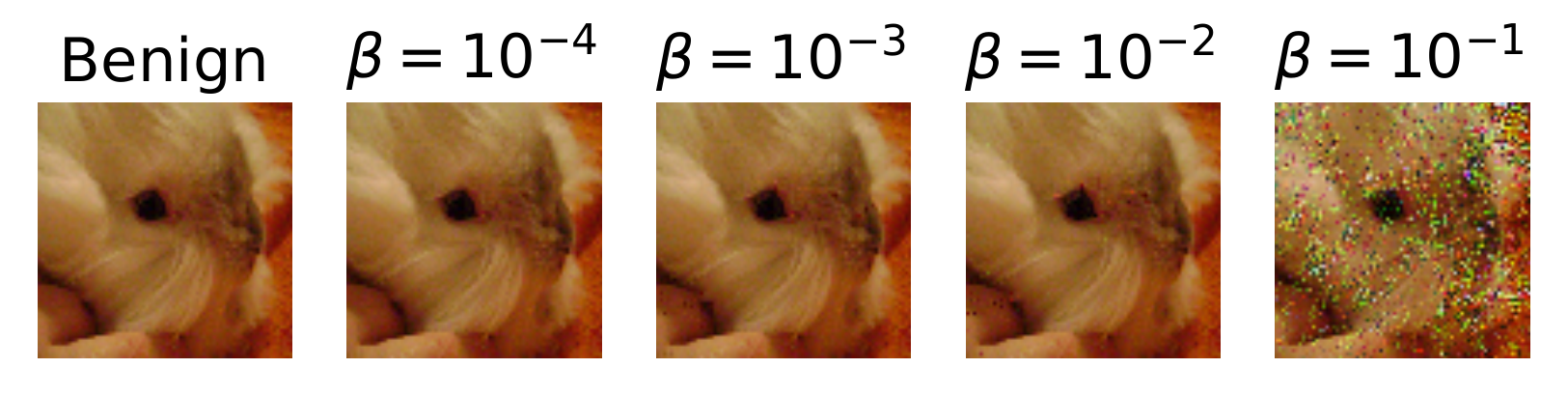}
\caption{ Adversarials generated by \tname-aware attacks.}\label{Fig.:sample_adaptive}
\end{figure}

Figure \ref{Fig.:sample_adaptive} shows two examples of adversarial inputs generated with different $\beta$ values. We can see that distortion artifacts are clearly visible for $\beta=10^{-1}$, and can be detected through other means. 

At high $\beta$ values, the attack is also less likely to succeed. For $\beta=10^{-1}$, only 58\% (VGG) and 31\% (ResNet) of examples can be converted into adversarials that defeat the unprotected baseline. \tname is still able to detect 81\% and 39\% of the VGG and ResNet ones, respectively. 

This shows that \tname is robust to defense-aware attacks that optimize for low $L_1$ distance.

\begin{figure}[htb]
  \centering
    \includegraphics[width=\linewidth]{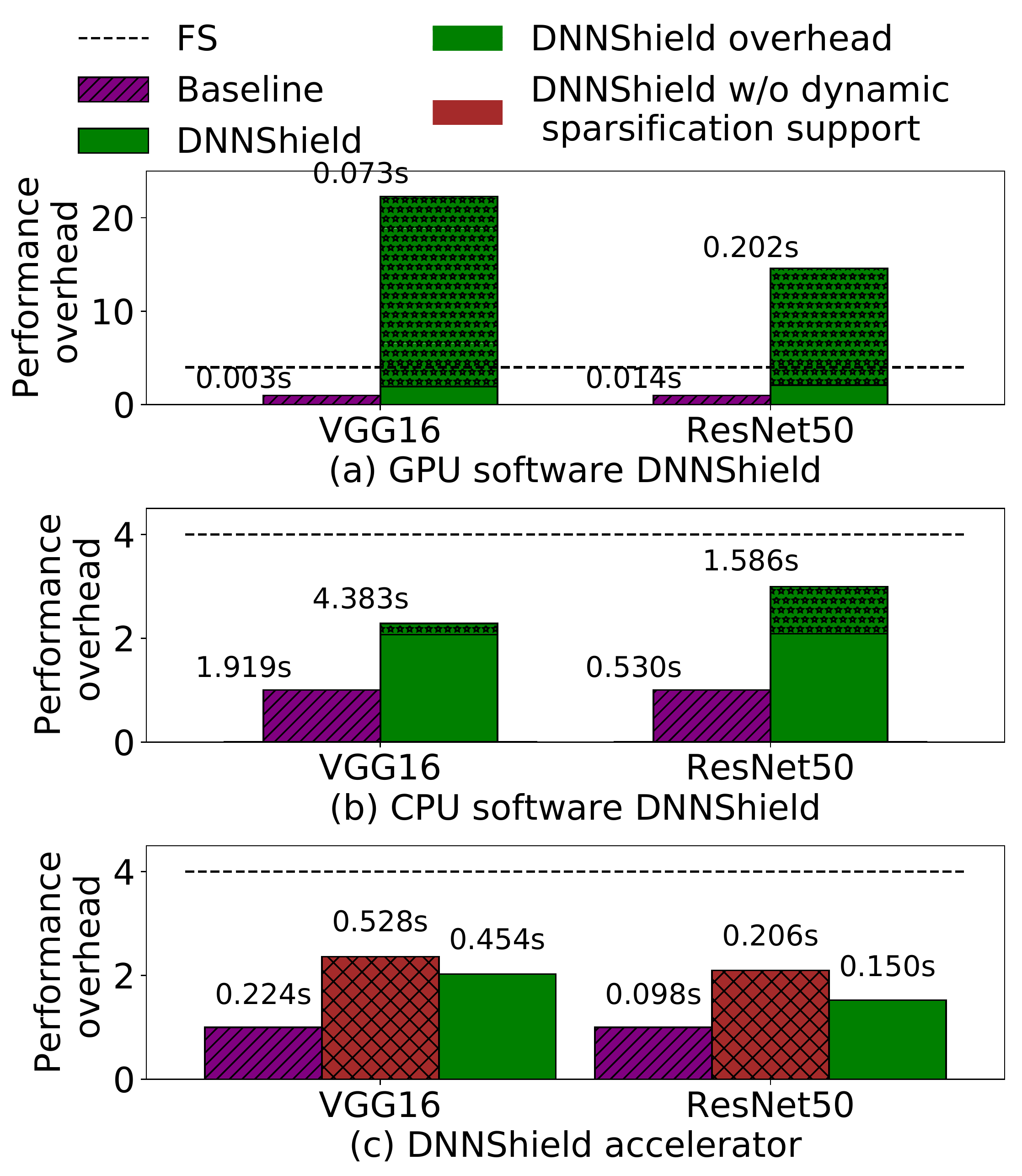}
  \caption{ \tname runtime on (a) GPU, (b) CPU and (c) \tname accelerator for VGG16 and ResNet50.}\label{Fig.:runtime}
\end{figure}

\subsection{Performance, Area and Power Overheads}

We next examine the performance, area and overheads of the \tname framework. Figure \ref{Fig.:runtime} shows the average normalized run time of  \tname on the GPU, CPU and \tname accelerator. The runtime overhead of software \tname on GPU is $15\times$ to $25\times$ higher than the baseline. This high overhead is primarily due to the random number generation function used by the dynamic sparsification algorithm -- which does not appear to be optimized on the GPU -- and is called when sparsifying each filter. This overhead is highlighted by the "\tname overhead", shown as a pattern in Figure \ref{Fig.:runtime}. 

In contrast, the overhead of the \tname accelerator implementation is much lower at $1.53 \times$ and $2 \times$ for ResNet50 and VGG16, respectively. Unlike the GPU, the \tname accelerator performance overhead is primarily due to re-execution of the approximate inference. While not trivial, the \tname performance overhead compares favorably with that of FS which exceeds $4\times$. For software-\tname on the CPU the overhead ranges from $2.43\times$ to $4.47\times$ which is again, higher that for \tname. In addition, the total runtime of the models on the CPU is dramatically longer than the FPGA. Very slow runtime of convolutional and FC layers on CPU dominate execution time. Hardware support for dynamic sparsification reduces overhead by 15\% and 30\% relative to the \tname without sparsification support. 

Table \ref{tab:area-power} summarizes the area and power overhead of the combined \tname hardware relative to the baseline CHaiDNN accelerator. We can see that the total overhead is low, with FPGA resource utilization increasing by at most 2.56\%. Power overhead is higher, but still small at $4.5\%$ dynamic. 

\begin{table}[htb]
\caption{FPGA resources and power overhead of \tname over baseline CHaiDNN accelerator.} 
\label{tab:area-power}
    \centering
    \scalebox{0.8}{
    \begin{tabular}{ |c|c|c|c|c|  }
        \hline
         Resource& Baseline &\tname&Overhead\%\\
         \hline
         BRAM   & 202.5    &204&   0.75\%\\
         DSP&   696  & 696   &0.0\\
         FF &112501 & 113630&  1.1\%\\
         LUT    &158060 & 159381&  0.8\%\\
         URAM&   80  & 80&0.00\\
         BUFG& 3  & 3   &0.00\\
         PLL& 1  & 1&0.00\\
         \hline
         \hline
         Power& Baseline &\tname&Overhead\%\\
         \hline
         Static   & 0.721W    &0.726&   0.6\\
         Dynamic&   5.567W  & 5.822W   &4.5\\
         \hline
     \end{tabular}
     }
\end{table}

\subsection{Sensitivity Studies} 
\label{sec:sensitivity}

\begin{figure}[htbp]
\centering
 \includegraphics[width=0.9\linewidth]{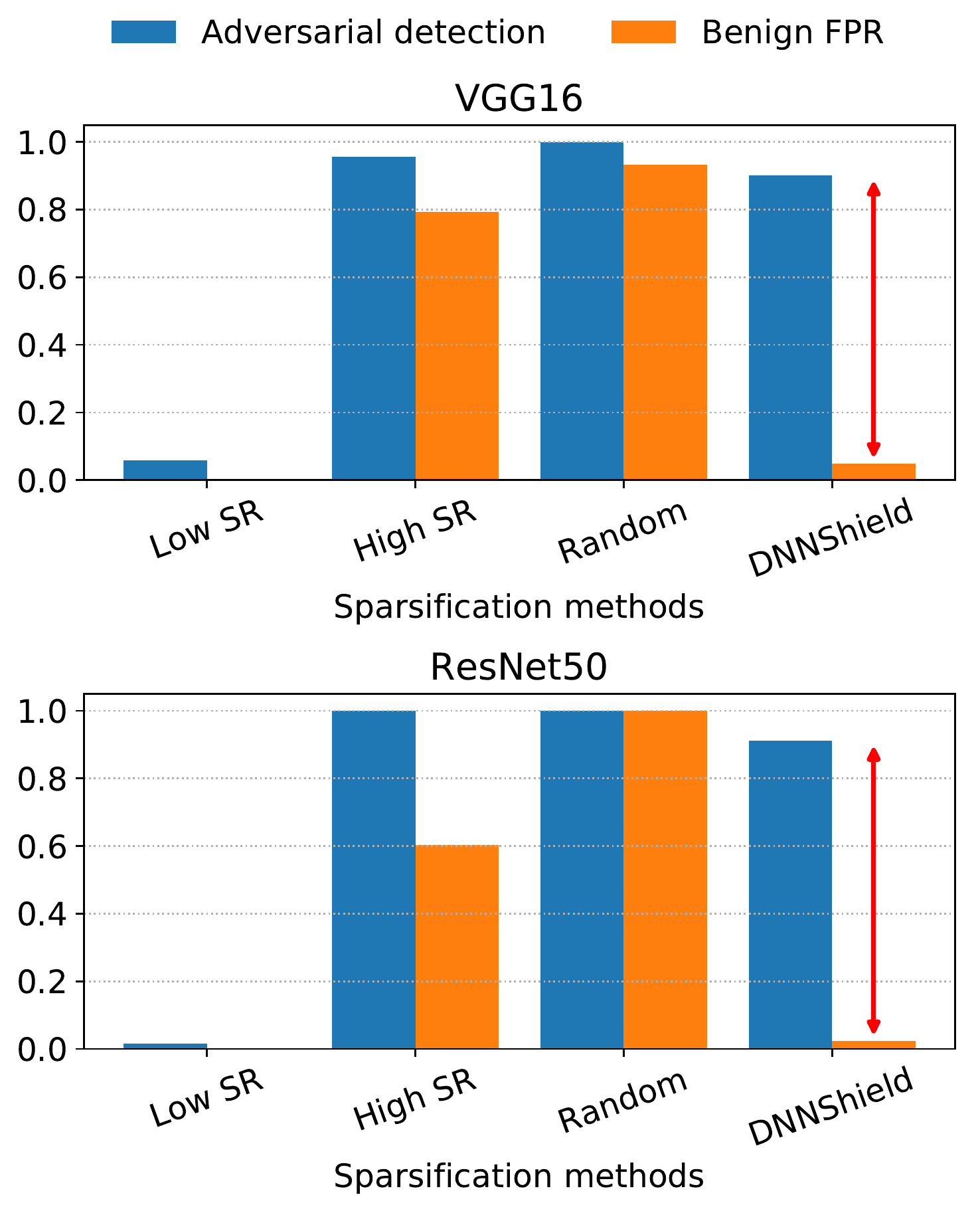}
\caption{ Adversarial detection and benign FPR with different sparsification approaches.} \label{Fig.:drop_method}
\end{figure}

The \tname design spans a broad design space that affects performance overhead for adversarial detection accuracy. 

\subsubsection{Sparsification Approaches}
We evaluate multiple approaches for dynamic sparsification. The naive approach of randomly dropping any weight subject to the sparsification rate (SR) results in, as Figure \ref{Fig.:drop_method} shows, a very high ($>90\%$) false positive rate (FPR) for benign inputs, indicating that random weight sparsification results cannot be used to discriminate adversarial inputs. This is because random sparsification can result in the dropping of large weight values, with large impact on classification output. To address this issue, in \tname we drop a random number of weights between 0 and SR from each filter, in ascending order of their values. This results in high adversarial detection, with low benign FPR. This is mostly due to the fact that dropping weights in ascending order enables more precise control over the approximation error. We also show that adapting the SR to classification confidence is very important. The High SR and Low SR experiments in Figure \ref{Fig.:drop_method} show the effects of weight dropping at fixed rates of up to $80\%$ and $20\%$ respectively. The Low SR is insufficient to achieve adversarial detection, while fixed $80\%$ results in very high benign FPR.

\begin{figure}[htbp]
\centering
 \includegraphics[width=0.8\linewidth]{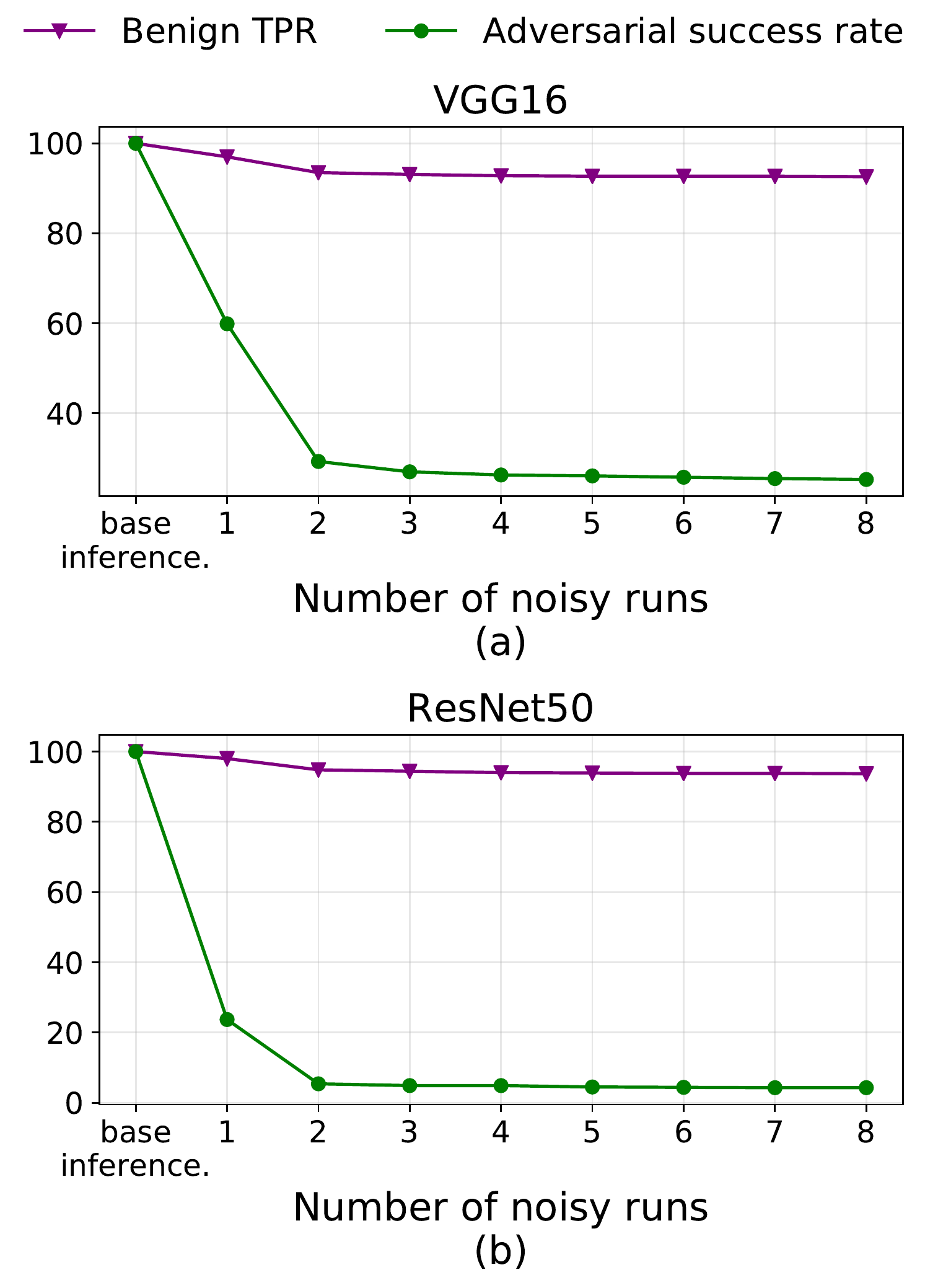}
\caption{ Adversarial attack success rate for multiple attacks as a function of the number of noisy runs in \tname.}\label{Fig.:acc_run}
\end{figure}

\subsubsection{Detection Convergence}

Figure \ref{Fig.:acc_run} shows the attack success rate as a function of the number of runs with inference approximation. More runs should ensure higher detection accuracy by generating more samples for the $L_1$ distance average. We can see that the attack success rate drops rapidly after 1-2 noisy runs, and remains mostly constant after that. This translates in \tname converging rapidly on a detection decision. A single noisy run is sufficient for $>$80\% of the benign inputs, and less than 10\% require more than 2.

\subsubsection{Detection Thresholds}

\begin{figure}[htbp]
\centering
 \includegraphics[width=0.8\linewidth]{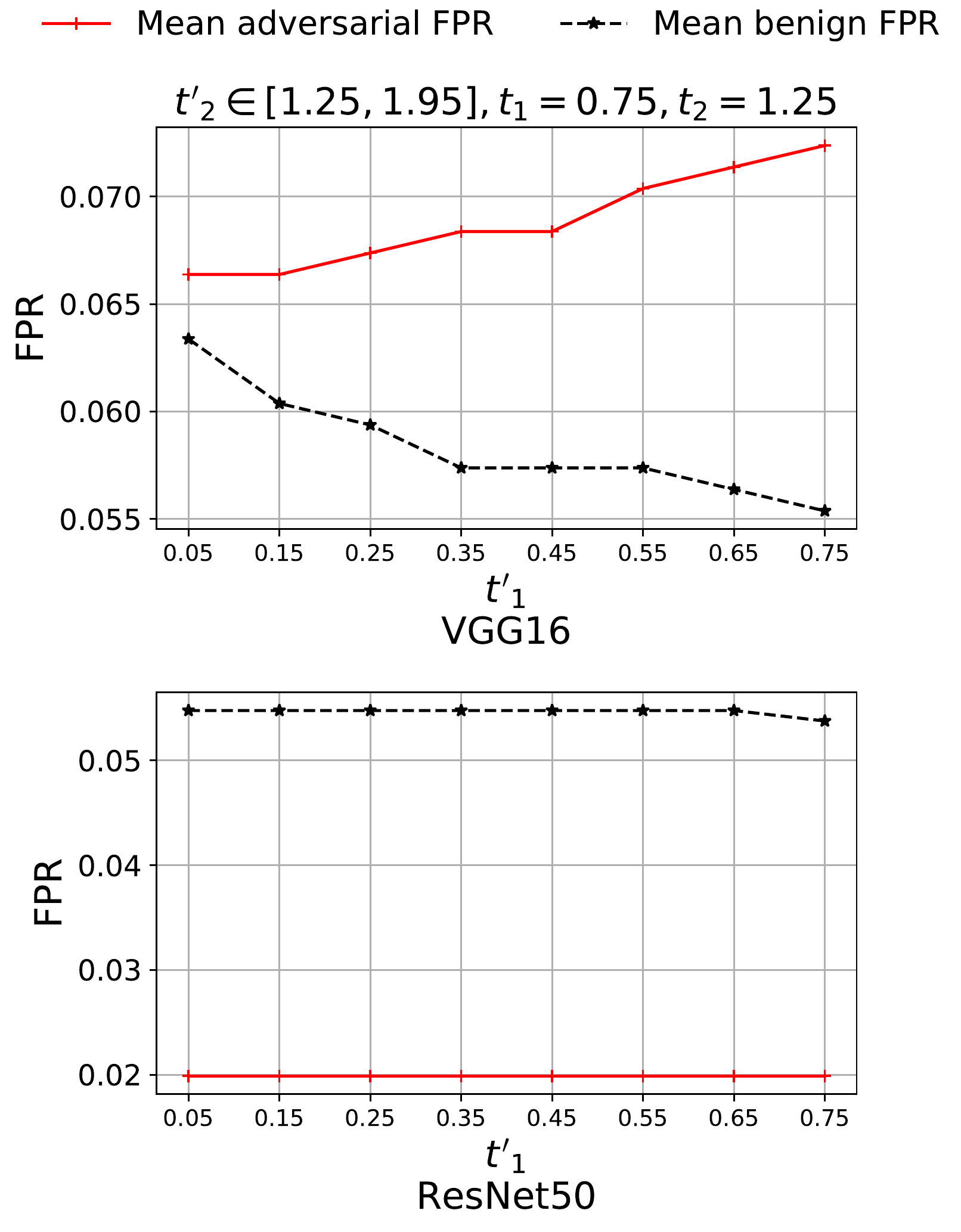}
\caption{ Benign  and adversarial FPR for different threshold values, left VGG16 and right ResNet50.}\label{Fig.:threshold_sen}
\end{figure}

Finally, we performed a sensitivity analysis on the threshold parameters used for adversarial detection.
To study the effect of detection thresholds, we varied $t'_1$ in the $[0.05,t_1]$ range in 0.1 increments. Then, for each value of $t'_1$ we varied $t'_2$ in the $[t_2,1.95]$ range and computed the average false positive rate (FPR).  Figure \ref{Fig.:threshold_sen} shows the average FPR for benign and adversarial inputs for different values of $t'_1$. ResNet50 exhibits a tighter distribution for $L_1$ distance under approximation 
and is therefore not sensitive to the threshold values. VGG16 on the other hand is more sensitive due to its wider distribution 
The threshold value allows a small tradeoff between FPRs for benigns vs. adversarials.  

\section{Related Work}
\label{sec:related}



Several other methods for designing robust neural networks to adversarial attacks have been proposed in the literature.  These methods typically fall into four broad categories \cite{akhtar2018threat}:

    
\noindent \textbf{Hardening the model}, also known as  \emph{adversarial training}. Recent works certified robust models by training models under Gaussian noise injection into the inputs \cite{cohen2019certified} or the model \cite{he2019parametric,lecuyer2019certified}. While these methods represent a systematic solution to adversarial attacks, they are limited to certain perturbation norms (e.g $L_2$) and do not scale for large datasets like ImageNet. Unlike model and input hardening approaches, \tname does \emph{not require any re-training} of the model and sacrifices little in model accuracy. \tname is designed to \emph{detect} adversarial examples post-training, during model inference.

\noindent \textbf{Hardening the test inputs}, also known as applying \emph{input transformations}, such as filtering or encoding the image. Input hardening methods require profiling to select appropriate parameters, such as Feature Squeezing \cite{xu2018feature} and Path Extraction \cite{qiu2019adversarial,9251936}. Xie et al. \cite{xie2018mitigating} propose to defend against adversarial examples by adding a randomization layer before the input to the classifier. \cite{9251936} showed that adversarial inputs tend to activate distinctive paths on neurons from those of benign inputs. They proposed hardware accelerated adversarial sample detection, which uses canary paths from offline profiling. In contrast, \tname does not require profiling.

\noindent \textbf{Adding a secondary, external network}  solely responsible for adversarial detection and with a separate training phase, such as NIC \cite{ma2019nic}. 
DNNGuard \cite{wang2020dnnguard} proposed an accelerator for such detection mechanism but has not evaluated a specific detection classifier. Secondary network detection-based methods are not as effective, and can be evaded by adaptive attacks \cite{carlini2017adversarial}. 

\noindent \textbf{Noise-based approaches} 
Prior work has similarly explored ways of discriminating adversarial inputs using noise. However, prior approaches have either proposed injecting noise into the input \cite{cohen2019certified, cao2017mitigating} -- with lower detection rate -- or into the model during training \cite{roth2019odds,he2019parametric,lecuyer2019certified,salman2020denoised}. However, the challenge with training-based approaches such as \cite{he2019parametric} is that the noise parameters tend to converge to zero as training progresses, making the noise injection progressively less effective over time \cite{jeddi2020learn2perturb}. While training-based approaches have enabled "certified robust" inputs, that certification is generally limited to a very narrow set of inputs.

\section{Conclusion}
\label{sec:conclusion}

In conclusion, this paper showed that dynamic and random sparsification of DNN models enables robust $>88\%$ adversarial detection across multiple strong attacks, for different image classifiers. We also demonstrated the importance of correlating approximation error to attack confidence, and showed robustness against defense-aware attacks.

\bibliographystyle{plain}
\bibliography{bibs/security,bibs/master.bib,bibs/rnn_hasi.bib}



\end{document}